\documentclass[epj,nopacs]{svjour}

\usepackage[stretch=10,shrink=10]{microtype}

\usepackage{amsmath}
\usepackage{amssymb}

\usepackage{graphicx}
\usepackage[bookmarks=true,colorlinks,linkcolor=blue,urlcolor=blue,citecolor=blue]{hyperref}
\usepackage[dvipsnames]{xcolor}
\usepackage{standalone}
\usepackage{float}
\usepackage{bm}
\usepackage{ulem}
\usepackage{cite}
\usepackage{dblfloatfix}

\makeatletter
\newcommand{\printfontsize}{\f@size pt}
\makeatother

\newcommand{\bpsi}{\bar{\psi}}

\definecolor{darkblue}{HTML}{004D6B}
\definecolor{darkred}{HTML}{8c1515}
\definecolor{darkgreen}{HTML}{006400}
\definecolor{crimson}{HTML}{1ABC9C}
\usepackage{hyperref}
\hypersetup{
	colorlinks=true,
	urlcolor=darkred,
	citecolor=darkblue,
	linkcolor=darkred,
	breaklinks
}
\newcommand{\veryshortarrow}[1][3pt]{\mathrel{%
   \hbox{\rule[\dimexpr\fontdimen22\textfont2-.2pt\relax]{#1}{.4pt}}%
   \mkern-7mu\hbox{\usefont{U}{lasy}{m}{n}\symbol{40}}}}
  
\title{%
Functional renormalization of spinless triangular-lattice fermions:\\ 
$N$-patch vs. truncated-unity scheme
}

\author{Nico Gneist${}^1$ \and Dominik Kiese${}^1$ \and Ravn Henkel${}^1$  \and Ronny Thomale${}^2$ \and Laura Classen${}^3$ \and Michael M. Scherer${}^4$}
\institute{Institute for Theoretical Physics, University of Cologne, 50937 Cologne, Germany \label{addr1}
\and 
Institute for Theoretical Physics, University of W\"urzburg, Am Hubland, D-97074 W\"urzburg, Germany
\and 
Max Planck Institute for Solid State Research, D-70569 Stuttgart, Germany \label{addr3}
\and
Institute for Theoretical Physics III, Ruhr-University Bochum, D-44801 Bochum, Germany \label{addr4}}

\date{\today}

\abstract{We study competing orders of spinless fermions in the triangular-lattice Hubbard model with nearest-neighbor interaction. We calculate the effective, momentum-resolved two-particle vertex in an unbiased way in terms of the functional renormalization group method and compare two different schemes for the momentum discretization, one based on dividing the Fermi surface into patches and one based on a channel decomposition. We study attractive and repulsive nearest-neighbor interaction and find a competition of pairing and charge instabilities. In the attractive case, a Pomeranchuk instability occurs at Van Hove filling and $f$-wave and $p$-wave pairing emerge when the filling is reduced. In the repulsive case, we obtain a charge density wave at Van Hove filling and extended $p$-wave pairing with reduced filling. The $p$-wave pairing solution is doubly degenerate and can realize chiral $p+ip$ superconductivity with different Chern numbers in the ground state. We discuss implications for strongly correlated spin-orbit coupled hexagonal electron systems such as moir\'e heterostructures. 
}

\begin{document}
\maketitle

\section{Introduction}

For decades the single-band Hubbard model has been the Standard Model of Correlated Electron Physics.
Not only has it been thought of as capturing essential features of the phase diagram of high-temperature superconductors and related materials, it has also served as a reference model for the development of quantum many-body methods~\cite{PhysRevX.11.011058}.

In terms of Fermi surface instabilities, the square lattice has been the dominating focus of theoretical research as it has been hosting the majority of quasi-two dimensional candidate materials for strongly correlated electron systems. More recently, however, the discovery of strongly-correlated states in moir\'e materials, i.e. systems based on few-layer stacks of two-dimensional materials such as graphene or transition metal dichalcogenides (TMD)~\cite{cao1,cao2,manzeli,wang2020correlated,tang2020}, has made a strong case for revisiting hexagonal lattice systems of correlated electrons from the viewpoint of state-of-the-art quantum many-body approaches~\cite{nandkishore,kiesel2012competing,PhysRevB.85.035414,PhysRevLett.110.126405,PhysRevLett.111.097001,PhysRevB.88.205121,PhysRevB.89.020509,PhysRevB.89.144501}.

Kagome, honeycomb, and triangular lattices all share the same hexagonal point group symmetry but differ in terms of Wyckoff positions taken by their respective lattice sites. The triangular lattice stands out as the local site symmetry matches that of the hexagonal point group symmetry. It has a high potential to offer exotic many-body states due an intricate interplay between frustration and correlations, see, e.g.,~\cite{PhysRevB.94.245145,PhysRevX.10.021042,PhysRevB.102.115136,PhysRevB.103.235132,chen2021quantum,PhysRevX.11.041013,zhu2020doped,PhysRevB.103.165138,scherer2021mathcal} for a recent series of studies on that matter.

In systems such as TMDs, a sizable spin-orbit coupling breaks the spin-rotation invariance. As a consequence, effective models for moir\'e TMDs often involve several spin-split bands~\cite{PhysRevLett.122.086402,zang2021,PhysRevResearch.2.033087} which need to be taken into account by adequate quantum many-body approaches. In an attempt to boil down moir\'e TMDs to its fermiological essence, this can thus lead to an effective model of spin-polarized interacting electrons (or spinless fermions). Note that in the absence of local Hubbard repulsion due to the removed spin degree of freedom, nearest-neighbor density-density interactions are the most elementary terms to consider, which we adopt for our paradigmatic toy model in the following.

A method that has been shown to be quite flexible when it comes to the description of competing instabilities of correlated-electron systems on various lattice geometries and for a broad range of fillings and interactions, is the functional renormalization group (FRG)~\cite{WETTERICH199390,salmhofer2001fermionic,dupuis2021nonperturbative}.
The FRG has been used in numerous studies to identify the leading Fermi-surface instabilities with all competing interaction channels being treated on equal footing~\cite{Metzner_review,platt2013functional}
Within the correlated-electron FRG different schemes have been employed for numerical implementations, most prominently the $N$-patch scheme, which divides the Brillouin zone into a number of $N$ patches with the representative momenta lying on the Fermi surface~\cite{PhysRevB.63.035109}.
The $N$-patch scheme allows for a relatively simple and straightforward numerical implementation, but becomes numerically expensive for high momentum resolution and also does not faithfully incorporate momentum conservation.
More recently, an alternative scheme -- the truncated-unity  scheme~\cite{lichtenstein2017high} (TUFRG) -- based on a decomposition of the different interaction channels~\cite{PhysRevB.79.195125} has been devised, which separates stronger and weaker momentum dependencies and therefore allows for a more efficient numerical evaluation at high momentum resolution.

In this work, we establish the correlated phase diagram of 
of spinless electrons on the triangular lattice in the presence of competing interaction channels around Van Hove filling.
To that end, we set up both, an $N$-Patch- and a TUFRG approach for correlated fermions without spin-SU(2) invariance.
We carefully study the convergence within both schemes and compare them to each other.
The motivation of our work is twofold:
\begin{enumerate}
 \item The FRG represents a very promising scheme for setting up sophisticated numerical implementations that can capture accurate multi-orbital/-band models for moir\'e TMDs. 
 Our results can then be used for future reference of such implementations.\medskip
\item The systematic quantitative comparison between the two FRG schemes provides guidance to the choice of transfer-momentum resolution and form-factor expansions in future TUFRG studies, which are likely to be more appropriate for a faithful description of more involved models due to numerical efficiency.
\end{enumerate}

\section{Model}

\begin{figure}[t!]
    \includegraphics[width=0.95\columnwidth]{ 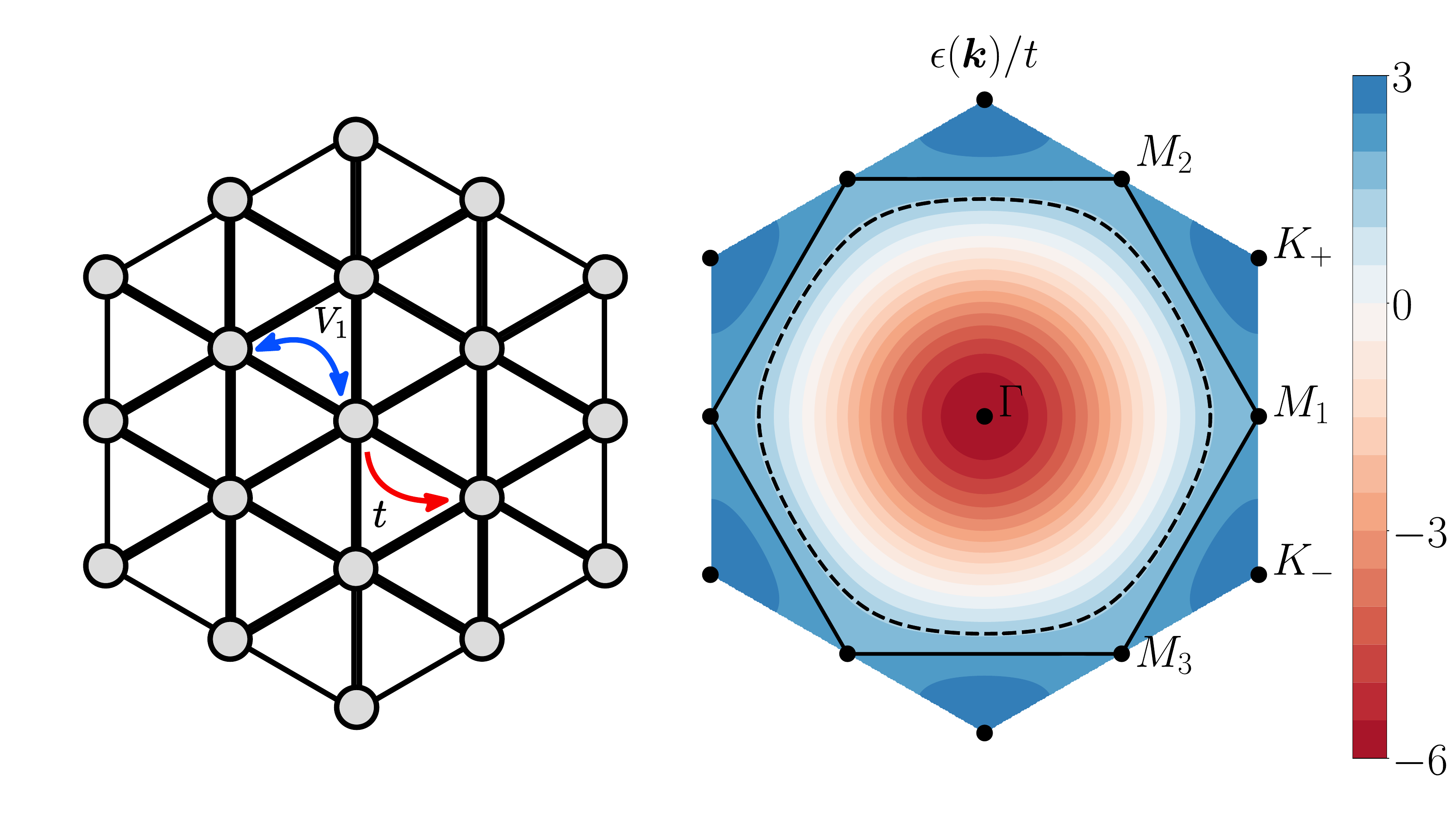}
    \caption{\textbf{Real-space lattice and dispersion in the BZ.} The solid line in the right panel shows the perfectly nested Fermi surface for $\mu / t = 2$ which corresponds to Van Hove filling. The dashed line shows the Fermi surface for $\mu / t = 1.4$.}
    \label{fig:model}
\end{figure}

We consider a tight-binding model for spinless fermions on the triangular lattice where we add a nearest-neighbor density-density interaction, reading
\begin{align}
    H =& -t\sum_{\langle ij\rangle}\left(c^\dagger_{i}c_{j}+ \mathrm{h.c.}\right)-\mu \sum_{i}n_{i}
    +V_1 \sum_{\langle ij\rangle }  n_{i}n_{j}\,.
    \label{eq:model}
\end{align}
Here the operator $c^{(\dagger)}_{i}$ annihilates (creates) a fermion on lattice site $i$, such that we allow for nearest-neighbor fermion hopping with rate $t$. 
The fermion density operator $n_i=c^\dagger_{i}c_{i}$ couples to the chemical potential $\mu$ to change the filling of the system and $V_1>0 \ (<0)$ is the strength of the repulsive (attractive) density interaction of neighboring fermions (see Fig.~\ref{fig:model}). 
We will study the effects of attractive and repulsive interactions for an extended range of fillings corresponding to $\mu$. 
The energy band of this model is given via a Fourier transform, yielding
\begin{align}\label{eq:dispersion}
   \xi(\bm{k})\!=\!-2t[\cos(k_x)\!+\!2\cos(k_x/2)\cos(\sqrt{3}k_y/2)]-\mu\,,
\end{align}
with wavevector $\bm{k}=(k_x,k_y)$. 
We note that at $\mu / t = 2$ the band dispersion features saddle points at the three inequivalent $\bm{M}$ points of the Brillouin zone (BZ) (Fig.~\ref{fig:model}), giving rise to a Van Hove singularity (VHS). 
Our investigations of the emergent many-body instabilities of the system will be carried out in the vicinity of the VHS, but also beyond.
%

\section{Fermionic functional renormalization group}

The fermionic functional renormalization group (FRG) \cite{WETTERICH199390,salmhofer2001fermionic} has been established as a versatile approach to treat strongly-correlated electrons without bias towards a specific mean-field channel~\cite{Metzner_review,platt2013functional}. 
It is rooted in the functional integral description of quantum many-body systems and it allows for the investigation of a broad range of models without specific limitations for their kinetic or interaction parameters.
Generally, the FRG acts as functional implementation of the Wilsonian renormalization-group (RG) idea, namely, one starts at an ultraviolet (UV) cutoff scale $\Lambda_{\text{UV}}$ and successively takes effects of fermionic fluctuations into account by approaching the infrared (IR) limit $\Lambda_{\text{IR}} = 0$.

While the FRG description of a selected model is at a formal level exact, one needs to decide for truncations of the description to derive a feasible numerical application from the general principles. 
In the situation of competing interactions, this truncation will mostly concentrate on the evolution of the two-particle vertex as an indicator for emerging Fermi-surface instabilities.
In the past, this has led to many successful applications of the method to strongly-correlated electron systems, for example, for models of spin-rotational invariant electrons on triangular and honeycomb lattices, see, e.g.,  ~\cite{PhysRevB.68.104510,honerkamp2008density,kiesel2012competing,PhysRevB.85.035414,PhysRevB.85.235408,PhysRevB.86.155415,PhysRevLett.110.126405,PhysRevLett.111.097001,PhysRevB.87.094521,PhysRevB.88.205121,PhysRevB.89.020509}.
In addition more specific models of these geometries have been investigated aiming at the description 
moiré materials ~\cite{PhysRevB.98.241407,PhysRevB.99.094521,PhysRevB.99.195120,scherer2021mathcal,gneist2022competing,2022arXiv220400648K}.

The FRG flow is realized by solving a system of coupled differential equations interpolating between the UV and the IR limit. 
In this work, we want to compare two specific computational schemes to track this FRG evolution of running couplings: (1)~the \textit{$N$-patch} scheme, which was one of the first well-established methods within the fermionic FRG framework, and (2)~the \textit{truncated-unit FRG}, a more recent approach which goes beyond the patching scheme and allows for a finer grained momentum resolution.

\subsection{Flow equations}

Our starting point is the action for a many-electron system
\begin{align}\label{eq:action}
	S[\bpsi, \psi]=-(\bpsi,G_0^{-1}\psi)+S_{\text{int}}[\bpsi,\psi]\,,
\end{align}
where $\bpsi,\psi$ are Grassmann-valued fields. Here, the quadratic term includes the free propagator $G_0(\omega,\bm{k})=1/(i\omega-\xi(\bm{k}))$ with Matsubara frequency $\omega$ and single-particle dispersion $\xi(\bm{k})$, and the bracket~$(.,.)$ denotes integrations over continuous and summations over discrete indices. The second term $S_{\text{int}}[\bpsi,\psi]$ in Eq.~\eqref{eq:action} is an interaction term, which can be read off directly from the interaction part of the microscopic Hamiltonian in Eq.~\eqref{eq:model}. With the help of the action $S$, we can define the Schwinger functional $\mathcal{G}[\bar\eta,\eta]=-\ln \int D \psi D\bar\psi \exp(-S[\bar\psi,\psi])\exp[(\bar\eta,\psi)+(\bar\psi,\eta)]$ and its Legendre transform - the effective action - $\Gamma[\bar\psi,\psi]=(\bar\eta,\psi)+(\bar\psi,\eta)+G[\bar\eta,\eta]$ with $\psi=-\partial G/\partial\bar\eta$ and $\bar \psi=\partial G/\partial\eta$, which generates the one-particle irreducible (1PI) correlation functions~\cite{negele2018quantum}.

The central step for setting up the renormalization group scheme amounts to regularizing the free propagator by an infrared cutoff $\Lambda$, such that $G_0(\omega,\bm{k})\to G_0^\Lambda(\omega,\bm{k})$. The cutoff implementation is, in some sense, arbitrary, as long as the ultraviolet ($\Lambda \to \infty$) and infrared limit ($\Lambda \to 0$) are smoothly connected. Here, we opt for implementing the temperature flow scheme introduced by Honerkamp and Salmhofer \cite{Honerkamp_Tflow} which is employed for both FRG implementations. For now, however, we will keep the discussion general and refer the reader to App.~\ref{app:regulator} for details on the $T$-flow.

Having regularized the bare propagator, the effective action $\Gamma[\bpsi,\psi]$ 
becomes scale dependent and its flow is governed by an exact differential equation~\cite{RevModPhys.84.299}, which reads
\begin{align} \label{eq:exRG}
	\frac{\partial}{\partial \Lambda}\Gamma^\Lambda\!=\!-(\bpsi,(\dot{G}^\Lambda_0)^{-1}\psi)\!-\!\frac{1}{2}\mathrm{Tr}\left(\!(\dot{\mathbf{G}}^\Lambda_0)^{-1}
	(\mathbf{\Gamma}^{(2)\Lambda})^{-1}
	\!\right)\,,
\end{align}
where $\mathbf{\Gamma^{(2)\Lambda}}=(\partial_{\bar\psi},\partial_\psi)^T (\partial_\psi,\partial_{\bar\psi})\Gamma^\Lambda $ is the matrix of second derivatives of $\Gamma^\Lambda$.
Here, the appearance of the matrix of second functional derivatives of the effective action $\mathbf{\Gamma}^{(2)}$ necessitates some truncation to derive a closed set of equations for the 1PI vertex functions. We employ a standard approximation scheme, which (1)~neglects self-energy insertions, such that undifferentiated fermion lines correspond to bare, unrenormalized propagators, (2)~sets external Matsubara frequency arguments to zero and, simultaneously, does not account for the frequency dependence of the two-particle vertex and (3)~truncates the three-particle vertex from the flow equations (an in-depth discussion of these approximations is reviewed in~\cite{Metzner_review}). As a result, we obtain flow equations for the static two-particle vertex $V(\bm{k}_1, \bm{k}_2, \bm{k}_3)$ (the fourth momentum is fixed by momentum conservation), which allow us to determine Fermi liquid instabilities in an unbiased way.

For spinless fermions, the flow equations read~\cite{PhysRevB.92.155137,PhysRevB.98.045142}
\begin{align}\label{eq:flowequation}
	\frac{d}{d\Lambda}V^\Lambda=\tau_\mathrm{pp}+\tau_\mathrm{ph,c}+\tau_\mathrm{ph,d}\,.
\end{align}
where
\begin{align}
    \tau_{\mathrm{pp}}=&-\frac{1}{2}\!\int_q  \frac{d}{d\Lambda}[G_0^\Lambda(i\omega, \bm{q}+\bm{k}_1+\bm{k}_2)G_0^\Lambda(-i\omega,-\bm{q})] \nonumber\\[3pt]
    & \quad\times V^{\Lambda}(\bm{k_1},\bm{k_2},\bm{q}+\bm{k_1}+\bm{k_2})\nonumber\\[3pt]
    &\quad\times  V^{\Lambda}(\bm{q}+\bm{k}_1+\bm{k}_2,-\bm{q},\bm{k}_3)\,,\label{eq::contribution1}
\end{align}
denotes the pairing or particle-particle channel
\begin{align}
    \tau_{\mathrm{ph,c}}=&-\!\int_q \frac{d}{d\Lambda}[G_0^\Lambda(i\omega, \bm{q}+\bm{k}_1-\bm{k}_4)G_0^\Lambda(i\omega,\bm{q})] \nonumber\\[3pt]
    & \quad\times V^{\Lambda}(\bm{k}_1,\bm{q},\bm{q}\!+\!\bm{k}_1\!-\bm{k}_4) \nonumber\\[3pt]
    &\quad\times V^{\Lambda}(\bm{q}+\bm{k}_1-\bm{k}_4,\bm{k}_2,\bm{k}_3)\,,\label{eq::contribution2}
\end{align}
the crossed particle-hole channel and
\begin{align}
    \tau_{\mathrm{ph,d}}=&+\!\int_q \frac{d}{d\Lambda}[G_0^\Lambda(i\omega, \bm{q}+\bm{k}_1-\bm{k}_3)G_0^\Lambda(i\omega,\bm{q})] \nonumber\\[3pt]
    & \quad\times V^{\Lambda}(\bm{k}_1,\bm{q},\bm{k}_3) \nonumber\\[3pt]
    &\quad\times V^{\Lambda}(\bm{q}\!+\!\bm{k}_1\!-\!\bm{k}_3,\bm{k}_2,\bm{q})\,,\label{eq::contribution3}
\end{align}
the direct particle-hole channel, respectively. Here the integral is defined as $\int_k= A_{\mathrm{BZ}}^{-1}T\int_{\mathrm{BZ}} d\bm{k} \sum_{i\omega}$ and $k=(\bm{k},\omega)$ where $A_{\mathrm{BZ}}$ is the area of the Brillouin zone. 

Integrating these equations starting with the bare coupling in the $\Lambda \to \infty$ limit, Fermi liquid instabilities are signified by singular contributions to $V$. We note that $V$ is a function of three momenta and it is therefore costly to compute. 
For this reason, we rely on further approximations for its momentum dependence, two of which are presented in the following.

\subsection{$N$-patch FRG}

\begin{figure}[t!]
    \centering
    \includegraphics[width=0.8\columnwidth]{ 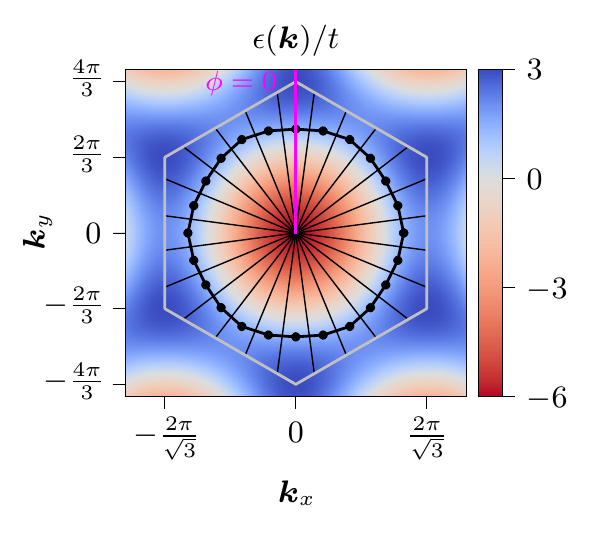}
    \vspace{-0.3cm}
    \caption{\textbf{Illustration of the $N$-patch FRG scheme} for $N = 24$ points on the Fermi surface (thick black line). The patches, indicated by thin black lines, range from the $\bm{\Gamma}$ point to the boundary of the first Brillouin zone (thick grey line). Our results are produced with $N=192$. The reference patch for the angular discretization is indicated by a thin magenta line.}
    \label{fig:patching_schematic}
\end{figure}

The first, well-established approximation of the momentum dependence assumes that the two-particle vertex is constant along elongated \textit{patches} in momentum space~\cite{Metzner_review}. 

To implement the patching scheme, we define a mapping $\pi: 1. \text{BZ} \to \mathbb{Z}^{N}_{\text{FS}}$, identifying momenta $\boldsymbol{k}$ in the first Brillouin zone with their nearest-neighbor $\pi({\boldsymbol{k}})$ in an angular discretization $\mathbb{Z}^{N}_{\text{FS}}$ of the Fermi surface, which consists of $N$ points, see Fig.~\ref{fig:patching_schematic}. This way, irrelevant couplings perpendicular to the Fermi surface are projected out and the vertex is fully determined by its value on the central patch points, which we place equidistantly. Note, that this treatment of the momentum dependence of the vertex spoils momentum conservation, since the fourth momentum $\boldsymbol{k}_4 = \pi({\boldsymbol{k}_1}) + \pi({\boldsymbol{k}_2}) - \pi({\boldsymbol{k}_3})$ of the projected vertex $V(\pi(\boldsymbol{k}_1), \pi(\boldsymbol{k}_2), \pi(\boldsymbol{k}_3))$ will in general not align with a patch point and therefore require an additional transformation with $\pi$. 

$N$-patch FRG calculations were successfully employed to track the flow of marginal couplings for prototypical model systems of high-$T_c$ superconductivity such as iron pnictides and cuprates, see, e.g., Refs. \cite{Metzner_review,platt2013functional,dupuis2021nonperturbative} and references therein.
This legitimates the method as a valid starting point to determine the leading instabilities around the Fermi surface fixed point.

In summary, the patching scheme describes the vertex with three projected momenta, i.e. $V(\pi(\boldsymbol{k}_1), \pi(\boldsymbol{k}_2), \pi(\boldsymbol{k}_3))$, such that for a selection of $N$ patches, the numerical cost will scale with $N^3$.
In this work, we implemented a resolution of the Fermi surface using $N = 192$ patches.

\subsection{Truncated-unity FRG}

The truncated-unity FRG (TUFRG)~\cite{lichtenstein2017high} allows for a high resolution of the full Brillouin zone, i.e. in contrast to the $N$-patch scheme, it is not restricted to the Fermi surface. 
Instead, one can chose arbitrary points of momenta to evaluate the flow equations.  
The derivation of the TUFRG approach is based on the fact that the singular behaviour of instabilities are mainly depending on the transfer momenta inside the loops in  Eqs.~\eqref{eq::contribution1}--\eqref{eq::contribution3} connecting the two vertices \cite{PhysRevB.79.195125}. Specifically, they are $\bm{k}_1+\bm{k}_2$ in $\tau_{pp}$, $\bm{k}_1-\bm{k}_4$ in $\tau_{ph,c}$ and $\bm{k}_1-\bm{k}_3$ in $\tau_{ph,d}$. Consequently, the interaction is re-parametrized into different channels such that each object is accounting for one of the transfer momenta. In practice, $V^{\Lambda}$ is decomposed as
\begin{align}
    V^{\Lambda}(\bm{k}_1,\bm{k}_2,\bm{k}_3,\bm{k}_4)=& \ V^{\Lambda,0}(\bm{k}_1,\bm{k}_2,\bm{k}_3,\bm{k}_4)  \nonumber\\[3pt] 
    &+\Phi^{\Lambda,P}(\bm{k}_1+\bm{k}_2;-\bm{k}_2,-\bm{k}_4)   \nonumber\\[3pt] &+\Phi^{\Lambda,C}(\bm{k}_1-\bm{k}_4;\bm{k}_4,\bm{k}_2)  \nonumber\\[3pt]
    &+\Phi^{\Lambda,D}(\bm{k}_1-\bm{k}_3;\bm{k}_3,\bm{k}_2)\,,
    \label{eq::decomposition}
\end{align}
where $V^{\Lambda,0}(\bm{k}_1,\bm{k}_2,\bm{k}_3,\bm{k}_4)$ accounts for the initial conditions of the model. The channels carry the important transfer momentum as first argument and each channel can be interpreted as representing a specific kind of interaction. The choice of these three channels was initially motivated by models of spinful fermions, where $P$ will represent a pairing interaction, and depending on spin combinations, $C$ and $D$ represent magnetic 
and density-density interactions. 
Since our model Eq.~\eqref{eq:model} is spinless, both channel $C$ and $D$ will eventually represent density-density interactions and this choice is therefore redundant
We keep this representation anyway such that a transfer of this method to a spinful model can be done in a transparent way.

\begin{figure}[t!]
    \includegraphics[width=\columnwidth]{ 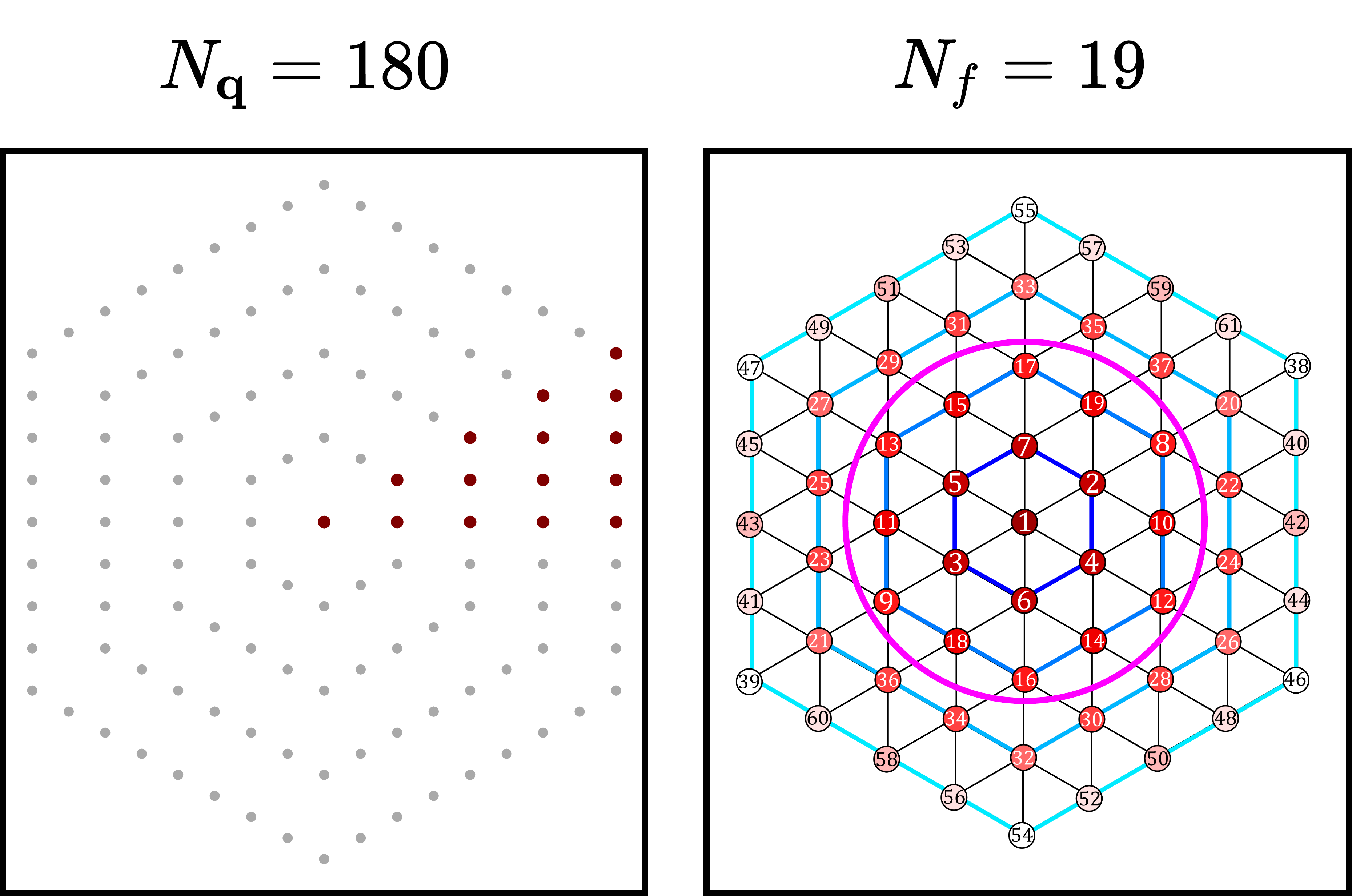}
    \caption{\textbf{Illustration of the TUFRG resolutions.} Left: for 
    the comparison with the patching scheme, $N_{\bm{q}}=180$ momentum points were chosen which are evenly spaced in the Brillouin zone. Only the contributions of the red points have to be calculated since the rest can be obtained by symmetry operations.  Right: The plane-wave form factors are $f_l(\bm{k})=\exp(i\bm{k}\bm{R}_l)$, where $\bm{R}_l$ are the real space vectors. Our results are produced with 
    $N_f=19$ (inside the magenta circle) unless stated otherwise. For more details see App.~\ref{app:momentaff}.}
    \label{fig:TUFRG_schematic}
\end{figure}

To relate the channels to the diagrams with the same important momentum, we define the flow equations
\begin{align}
    \frac{d}{d\Lambda} \Phi^{P}(\bm{k}_1+\bm{k}_2;-\bm{k}_2,-\bm{k}_4) &= \tau_{\mathrm{pp}}(\bm{k}_1,\bm{k}_2,\bm{k}_3,\bm{k}_4)\label{eq::diff3a}\,,  \\
    \frac{d}{d\Lambda} \Phi^{C}(\bm{k}_1 - \bm{k}_4;\bm{k}_4,\bm{k}_2) &= \tau_{\mathrm{ph,c}}(\bm{k}_1,\bm{k}_2,\bm{k}_3,\bm{k}_4)\,,  \label{eq::diff3b} \\
    \frac{d}{d\Lambda} \Phi^{D}(\bm{k}_1-\bm{k}_3;\bm{k}_3,\bm{k}_2) &= \tau_{\mathrm{ph,d}}(\bm{k}_1,\bm{k}_2,\bm{k}_3,\bm{k}_4)\label{eq::diff3c}\,,
\end{align}
where $\Lambda$ was dropped for brevity. Since the the last two momenta of the channels are deemed as less important, we will expand them in form-factors:
\begin{equation}
    \Phi^{X}(\bm{q},\bm{k},\bm{k}') = \sum_{l,l'} X^{l,l'}(\bm{q}) f_l(\bm{k}) f_{l'}^{*}(\bm{k}')\,\label{eq::formfactorexpansion}
\end{equation}
with $X \in \{P,C,D\}$. This expansion can be imposed as long as the form-factors are forming a unity:
\begin{align}
    A_{\mathrm{BZ}}^{-1}\sum_{l} f_{l}^{*}(\bm{p})f_{l}(\bm{k})&= \delta(\bm{p}-\bm{k})\,, \label{eq::unity1}\\
    A_{\mathrm{BZ}}^{-1}\int d\bm{k} f_{l}^{*}(\bm{k})f_{l'}(\bm{k})&= \delta_{l,l'}\,.
    \label{eq::unity}
\end{align}
The channel decomposition and the unity of the form-factors can now be used to reformulate the initial flow equations into a form which offers a computational advantage.

In the TUFRG approach we derive flow equations for $P^{l,l'}(\bm{q}),C^{l,l'}(\bm{q}),D^{l,l'}(\bm{q})$ by taking the derivative $\frac{d}{d\Lambda}$ and inserting form-factor resolved unities on the right hand side of Eqs.~\eqref{eq::contribution1}--\eqref{eq::contribution3} between the vertices and the loops, eventually leading to separating the three objects momentum-wise while connecting them in terms of form-factors. 
The sum of the form factors introduced with the unity Eq.~\eqref{eq::unity} can then be truncated safely to gain a numerical  advantage. 
The final form of the TUFRG flow equations reads
\begin{align}
    \frac{d}{d\Lambda}P^{l,l'}(\bm{q}) &= +\frac{1}{2}\sum_{l_1,l_2}  V^{P}(\bm{q})_{l,l_1} \dot{B}(\bm{q})_{l_1,l_2}^{(-)}V^{P}(\bm{q})_{l_2,l'} \,, \label{eq::ffflowP} \\
    \frac{d}{d\Lambda}C^{l,l'}(\bm{q}) &=  +\sum_{l_1,l_2}  V^{C}(\bm{q})_{l,l_1} \dot{B}(\bm{q})_{l_1,l_2}^{(+)}V^{C}(\bm{q})_{l_2,l'}\,, \label{eq::ffflowC}\\
    \frac{d}{d\Lambda}D^{l,l'}(\bm{q}) &=  -\sum_{l_1,l_2}  V^{D}(\bm{q})_{l,l_1} \dot{B}(\bm{q})_{l_1,l_2}^{(+)}V^{D}(\bm{q})_{l_2,l'}\,,\label{eq::ffflowD}
\end{align}
for details of the objects see App.~\ref{app:details}.

\begin{figure}[t!]
    \includegraphics[width=\columnwidth]{ 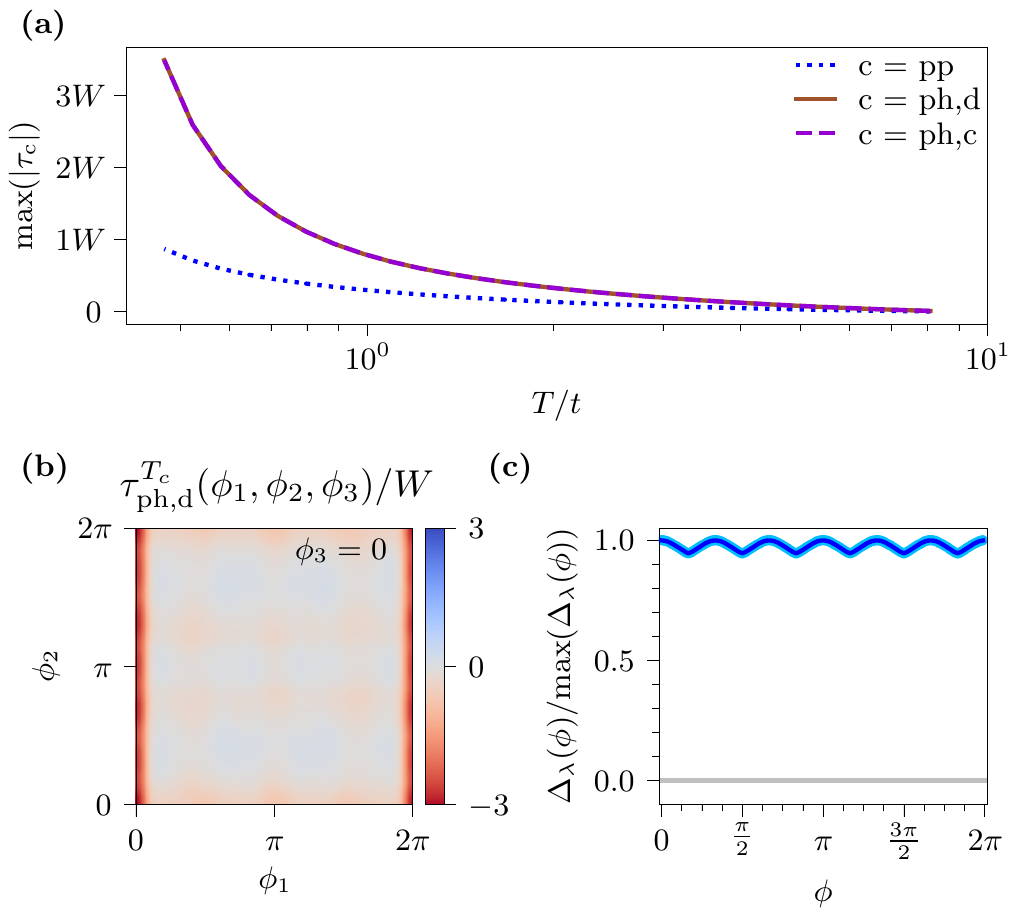}
    \caption{\textbf{Patching results for $V_1 / t = -1$ and $\mu / t = 2$.} The flow of the largest couplings in each channel is plotted in (a) and indicates an instability in the particle-hole channels. Solving the respective gap equation using the renormalized vertex in (b), yields the largest eigenvalue for transfer momentum $\bm{q} = 0$. The so-determined order parameter, indicated by light blue dots in (c), has an extended $s$-wave symmetry and transforms in the $A_1$ irrep. with both first and second neighbor harmonics (a fit to the numerical data is plotted as a dark blue line).}
    \label{fig:attractive20_patching}
\end{figure}

The flow equations now scale with $N_{\bm{q}}\times N_f^2$, where $N_{\bm{q}}$ is the number of momenta $\bm{q}$ which discretize the Brillouin zone and $N_f$ is the number of chosen form-factors, see Fig.~\ref{fig:TUFRG_schematic}. 
In practice one has to choose much less form-factors than patches in the patching scheme. 
Therefore, we gain a numerical advantage over the scaling of the $N$-patch scheme ($\sim N^3$) and the freedom to choose a larger number of momenta $N_{\bm{q}}$ in the Brillouin zone. 
In this work,  we use $N_{\bm{q}} = 180$ and $N_f = 19$ for comparison with $192$ patches in the other approach. To discuss single points in the phase diagram we use $N_{\bm{q}} = 540$ and $N_f = 19$. 
In the convergence checks we go up to $N_{\bm{q}}=792$ and $N_f=61$.
For details about the choice of momenta and form factors, see App.~\ref{app:momentaff}. 

\subsection{Linearized gap equation}
\label{sub:gaps}

To obtain the gap function $\Delta(\bm{k})$ for the superconducting instabilities encountered during the FRG flow, we utilize standard BCS theory \cite{RevModPhys.63.239}, that is, we perform a mean-field decoupling in the superconducting channel and derive a self consistent gap equation for $\Delta(\bm{k})$. Close to the critical temperature, where the gap is presumably small, the gap equation can be linearized and resembles an eigenvalue equation, which reads
\begin{align}
    \Delta(\bm{k}) = -\frac{1}{N} \sum_{\bm{k}'} V_{\text{BCS}}(\bm{k},\bm{k}') \frac{\Delta(\bm{k}')}{2\xi_{\bm{k}'}} \mathrm{tanh}\left(\frac{\xi_{\bm{k}'}}{2T_c} \right )\,.
    \label{eq:BCS_gap}
\end{align}
The only input required to solve Eq.~\eqref{eq:BCS_gap} and determine the leading contributions to the gap function as the eigenvectors with the largest negative eigenvalues, is then given by the pairing potential $V_{\text{BCS}}(\bm{k},\bm{k}') = V(\bm{k}, \bm{-k}, \bm{k}', -\bm{k}')$.

For the patching approach, we rewrite the right hand side of Eq.~\eqref{eq:BCS_gap} as an integral over a small energy shell $-\epsilon_c \leq \xi_{\bm{k}} \leq \epsilon_c \ll \epsilon_{\text{FS}}$ around the Fermi surface, where the most dominant contribution to the momentum sum stems from. The gap equation thus becomes
\begin{align}
    \Delta(\bm{k}) \approx & - \left[\int_{-\epsilon_c}^{\epsilon_c} d\xi \ \frac{1}{2\xi} \mathrm{tanh}\left(\frac{\xi}{2T_c} \right ) \right]\nonumber\\
    &\quad\times \langle V_{\text{BCS}}(\bm{k},\bm{k}') \Delta(\bm{k}') \rangle_{\bm{k}' \in \text{FS}} \,,
\end{align}
where the integral evaluates to
\begin{align}
    \int_{-\epsilon_c}^{\epsilon_c} d\xi \ \frac{1}{2\xi} \mathrm{tanh}\left(\frac{\xi}{2T_c} \right ) \approx \text{ln}\left(1.13 \frac{\epsilon_c}{T_c} \right) \,.
    \label{eq:FS_gap}
\end{align}
Finally, substituting $V_{\text{BCS}}(\bm{k}, \bm{k}') = \tau^{T_c}_{\text{pp}}(\bm{k}, \bm{-k}, \bm{k}', -\bm{k}')$ in Eq.~\eqref{eq:FS_gap} allows to straightforwardly obtain $\Delta(\bm{k})$ on the Fermi surface within the patching approach.

If we work with the TUFRG approach instead, we can restore the pairing interaction straightforwardly by calculating the pairing interaction from the $P$ channel, which is just given by the form-factor expansion in Eq.~\eqref{eq::formfactorexpansion}:
\begin{align}\label{eq:pairgap}
     \Phi^{P}(\bm{q},\bm{k},\bm{k}') = \sum_{l,l'} P^{l,l'}(\bm{q}) f_l(\bm{k}) f_{l'}^{*}(\bm{k}').
\end{align}
Since the divergence has a sharp peak at $q=0$, we set the superconducting pairing interaction as:
\begin{align}
\Phi^{P}(\bm{q}=0,\bm{k},\bm{k}'):=\Phi^{P}(\bm{k},\bm{k}')\,,
\end{align}
and identify $V_{\text{BCS}}(\bm{k},\bm{k}')=\Phi^{P}(\bm{k},\bm{k}')$. Thereafter, the gap function is obtained by diagonalization of the $N_{\bm{q}} \times N_{\bm{q}}$ matrix $\Phi^{P}(\bm{k},\bm{k}')$.

Note, that while we have focused on pairing instabilities for the sake of brevity, one can generalize the discussion above directly to instabilities in the particle-hole channels by performing the respective mean-field decoupling and deriving a gap equation with an appropriate density instead of a pairing potential. 


\section{Attractive case $V_1 < 0$}

\begin{figure}[t!]
    \includegraphics[width=\columnwidth]{ 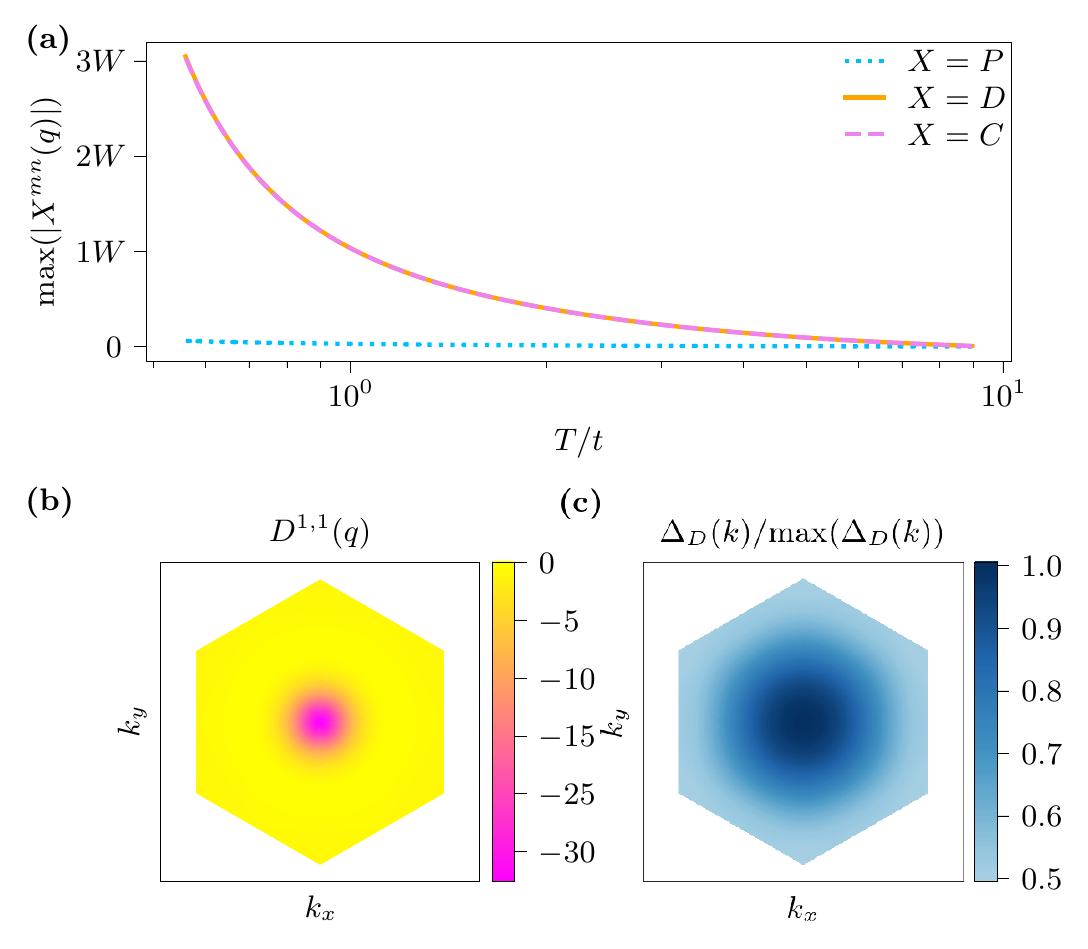}
    \caption{\textbf{TUFRG results for $V_1 / t = -1$ and $\mu / t = 2$}. Tracking the evolution of the $P,C,D$ channels, we can compare the maximal value of the respective vertices and detect a divergence in the $C/D$ channel, see (a). Moreover we notice the expected alignment of the $D$ and $C$ channel due to symmetry. The momentum resolved on-site vertex $D^{1,1}(q)$ in (b) peaks at the $\bm{\Gamma}$ point, indicating the possibility of a Pomeranchuck instability. The reconstructed order parameter $\Delta_D(k)$ of the $D$ channel is in the $A_1$ irrep., see (c). We use $N_{\bm{q}}=540$, $N_f=19$.}
    \label{fig:attractivePanel20}
\end{figure}

We first investigate the case of attractive interactions $V_1 < 0$ at and away from Van Hove filling $\mu / t = 2$. To that end, we apply both, the $N$-patch and the TUFRG scheme, and work out the qualitative and quantitative differences between these approaches. In order to generate a common starting point we initialize both methods as follows: the RG flow starts at
\begin{align}
    T_{\mathrm{UV}}  = W \,,
\end{align}
where $W = 9t$ is the bandwidth of the model. The respective flow equations are integrated down to the infrared, which we numerically define by $T_{\mathrm{IR}} / t = 10^{-5}$. If one of the channels diverges, signified by its maximum exceeding $3W$, the integration is terminated preemptively. 
As initial value for the vertex, we set $V^W(\pi(\boldsymbol{k}_1), \pi(\boldsymbol{k}_2), \pi(\boldsymbol{k}_3))=V_1$ in the patching scheme, and $V^{W}(\bm{k}_1,\bm{k}_2,\bm{k}_3,\bm{k}_4)=V_1$, $\Phi^{W,X}=0$ in the TUFRG.

\subsection{Pomeranchuk instability at Van Hove filling}

Tracking the evolution of the attractive case under the RG flow both employed approaches eventually detect a divergence of the particle-hole channels between $T / t = 1$ and $T / t = 0.1$, see Figs.~\ref{fig:attractive20_patching} and~\ref{fig:attractivePanel20}. 
Due to crossing symmetry, which relates the direct and crossed particle-hole contributions, the flows of the respective maxima align and we, thus, reduce our discussion to $\tau_{\mathrm{ph, d}}$ for brevity.

In the patching scheme, we find the most singular eigenvalue to emerge from the linearized gap equation (see App.~\ref{app:gaps} for further details) with transfer momentum $\bm{q} = \bm{k_1} - \bm{k_3} = 0$, corresponding to a Pomeranchuk instability~\cite{PhysRevX.6.041045}. The respective order parameter $\langle \bpsi_{\bm{k}} \psi_{\bm{k}} \rangle$ (see (c) in Fig.~\ref{fig:attractive20_patching}) is found to live in the $A_1$ irreducible representation (irrep.) of $C_{6v}$ with an extended $s$-wave form factor including nearest and second-nearest neighbors. The momentum modulation, induced by the second neighbor harmonic, is, however, quite weak to the constant offset presented by the nearest-neighbor $A_1$ basis function.

In the TUFRG scheme, the instability almost exclusively affects the onsite-component $D^{1,1}(q)$ of the direct particle-hole channel, with a pronounced peak at the $\bm{\Gamma}$ point (see Fig.~\ref{fig:attractivePanel20} (b)) and in agreement with the patching results. The reconstructed order parameter $\Delta_{D}$ (see (c) in Fig.~\ref{fig:attractivePanel20}) likewise transforms in the $A_1$ irrep., including a momentum modulation on the Fermi line. Note that, due to this modulation being weak compared to the nearest-neighbor $A_1$ contribution, this is rather difficult to see from the colormap in Fig.~\ref{fig:attractivePanel20}.

\begin{figure}[t!]
    \includegraphics[width=\columnwidth]{ 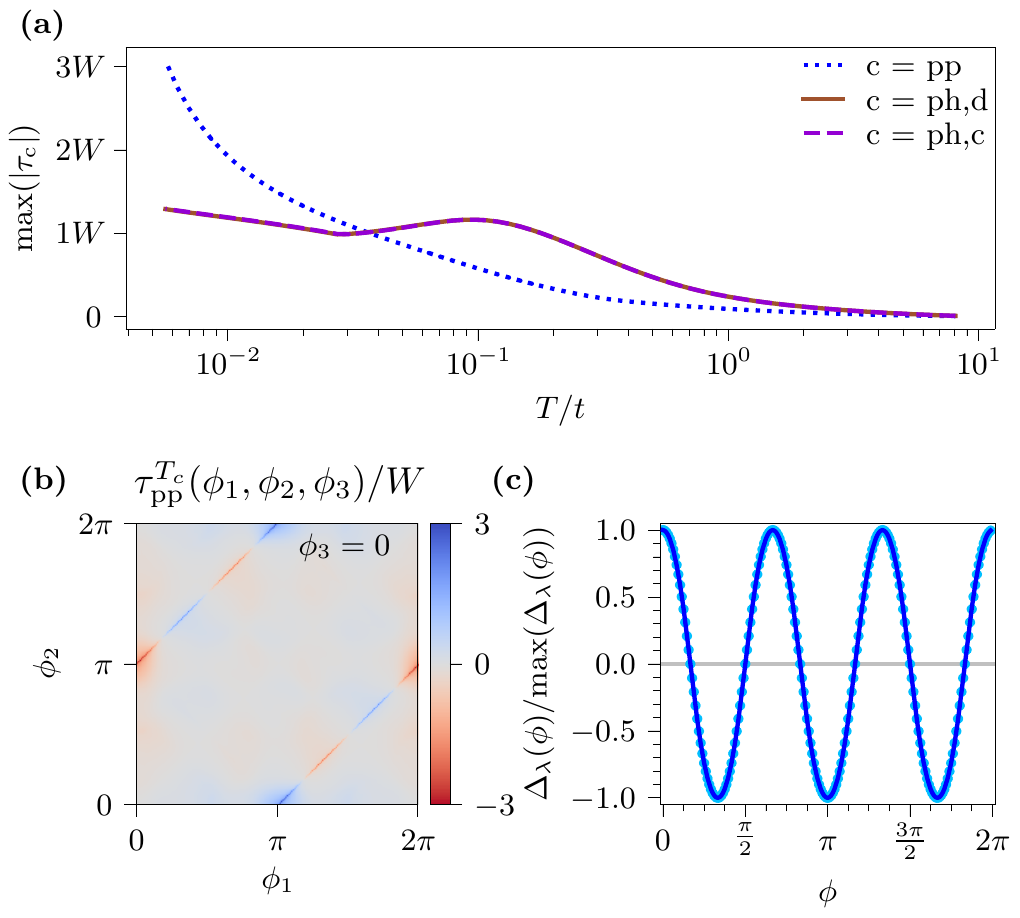}
    \caption{\textbf{Patching results for $V_1 / t = -0.6$ and $\mu / t = 1.8$.} Here, the flows of the channel maxima (see (a)) signifies a pairing instability. The superconducting gap, extracted from the renormalized vertex in (b), has $f$-wave symmetry (light blue dots) and can be fitted by the nearest-neighbor harmonic of the $B_1$ irrep. (dark blue line).}
    \label{fig:attractive18_patching}
\end{figure}

\subsection{Superconductivity below Van Hove filling}

For fillings $\mu / t < 2$, the Fermi surface is deformed and at some point the Pomeranchuck instability is overruled by a superconducting instability.
We observe that, depending on the combination of chemical potential and interaction strength, both employed FRG schemes consistently predict two different kinds of superconductivity with $\bm{q} = 0$ for an extended range of fillings.

\begin{figure}[t!]
    \includegraphics[width=\columnwidth]{ 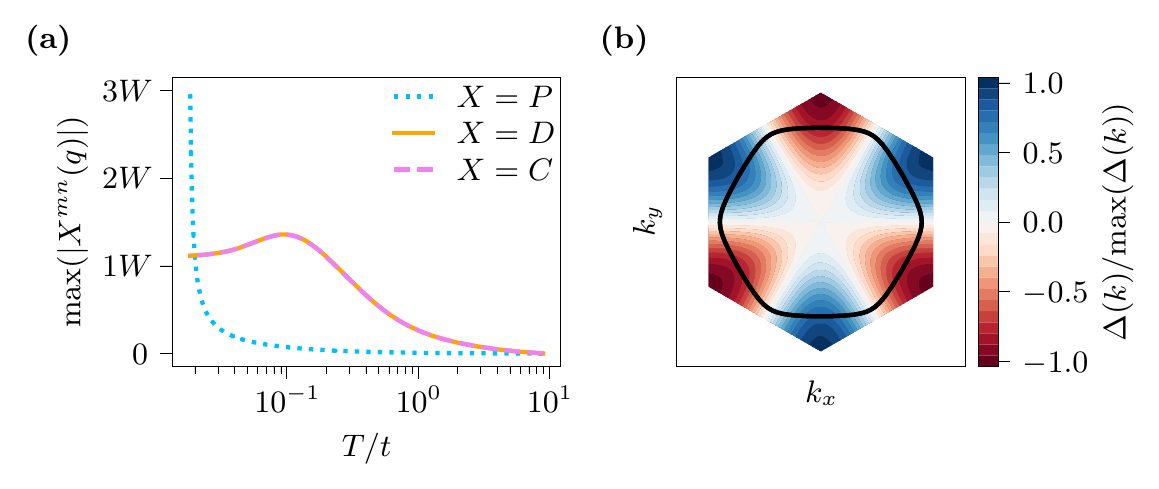}
    \caption{\textbf{TUFRG results for $V_1 / t = -0.6$ and $\mu / t = 1.8$.} Tracking similar to Fig.\ref{fig:attractivePanel20} we can now find a divergence of the $P$ channel away from Van Hove filling, indicating the emergence of superconductive instability (see (a)).  The reconstructed leading gap $\Delta(k)$ of this instability (see (b)) depicts a function in the $B_1$ irrep. of $C_{6v}$. The black line represents the Fermi surface, featuring 6 zero crossings. We use $N_{\bm{q}}=540$, $N_f=19$.}
    \label{fig:attractivePanel18}
\end{figure}

More specifically, moving away from $\mu/t=2$ towards smaller values, the Pomeranchuck instability will at first be replaced by a region of $f$-wave superconductivity.
Using the FRG data, we can reconstruct a gap function as detailed in Sec.~\ref{sub:gaps}.
Indeed, we find that the gap function belongs to the one-dimensional $B_1$ irrep. of $C_{6v}$ (see Figs.~\ref{fig:attractive18_patching} and \ref{fig:attractivePanel18}).
The size of the filling range where the $f$-wave superconductivity instability occurs grows for decreasing interaction strength $|V_1|$. 
In the case of $V_1=-0.4$ this type of superconductivity is even the only one which persists. 
Notably, in the case of $V_1=-1.0$, where the region is the smallest, the $N$-patch FRG scheme does not detect $f$-wave at all while TUFRG still resolves a small domain of this instability.

\begin{figure}[t!]
    \includegraphics[width=\columnwidth]{ 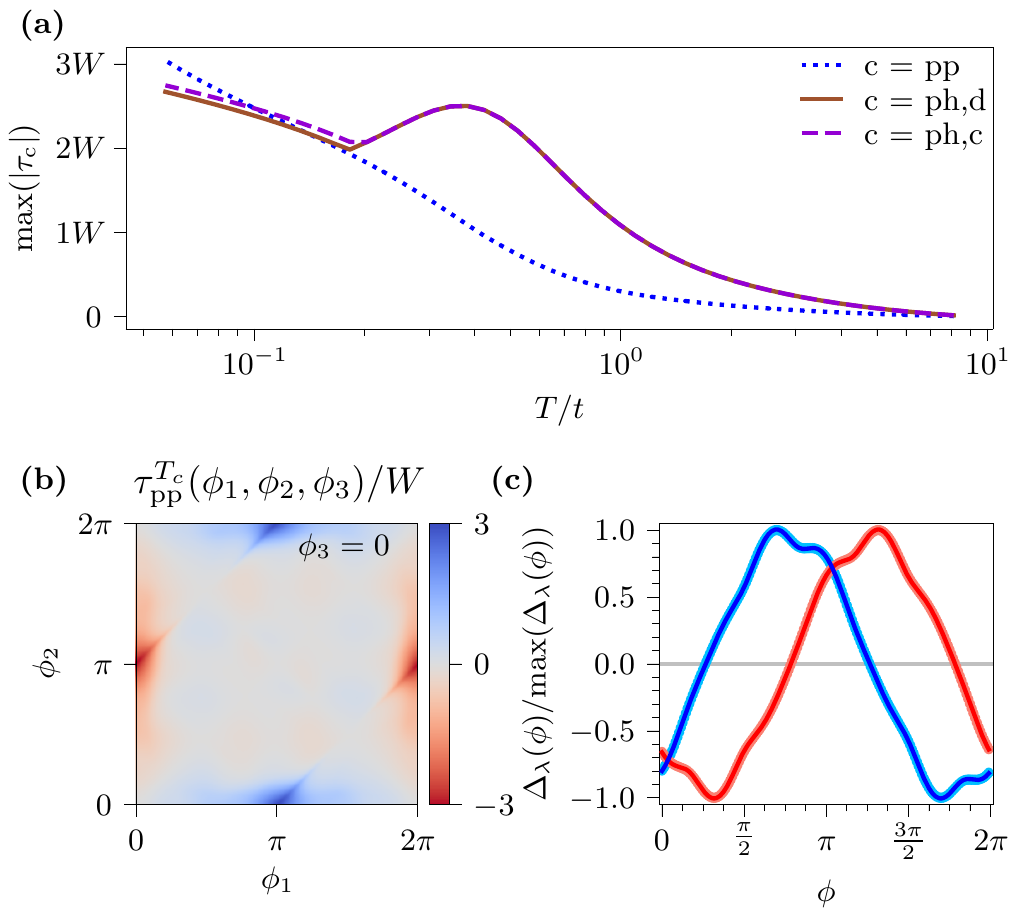}
    \caption{\textbf{Patching results for $V_1 / t = -1$ and $\mu / t = 1.2$.} Similar to Fig.~\ref{fig:attractive18_patching}, the vertex flows, plotted in (a), hint towards a superconducting instability. The respective gap equation, which requires the renormalized vertex from (b) as input, has a two-fold degenerate leading eigenvalue. The respective eigenvectors (superconducting gaps), displayed as light blue (light red) dots in (c), have $p$-wave symmetry and are well described by the nearest-neighbor lattice harmonics of the $E_1$ representation of $C_{6v}$, which we indicate by a dark blue (dark red) line.}
    \label{fig:attractive12_patching}
\end{figure}

\begin{figure}[t!]
    \includegraphics[width=\columnwidth]{ 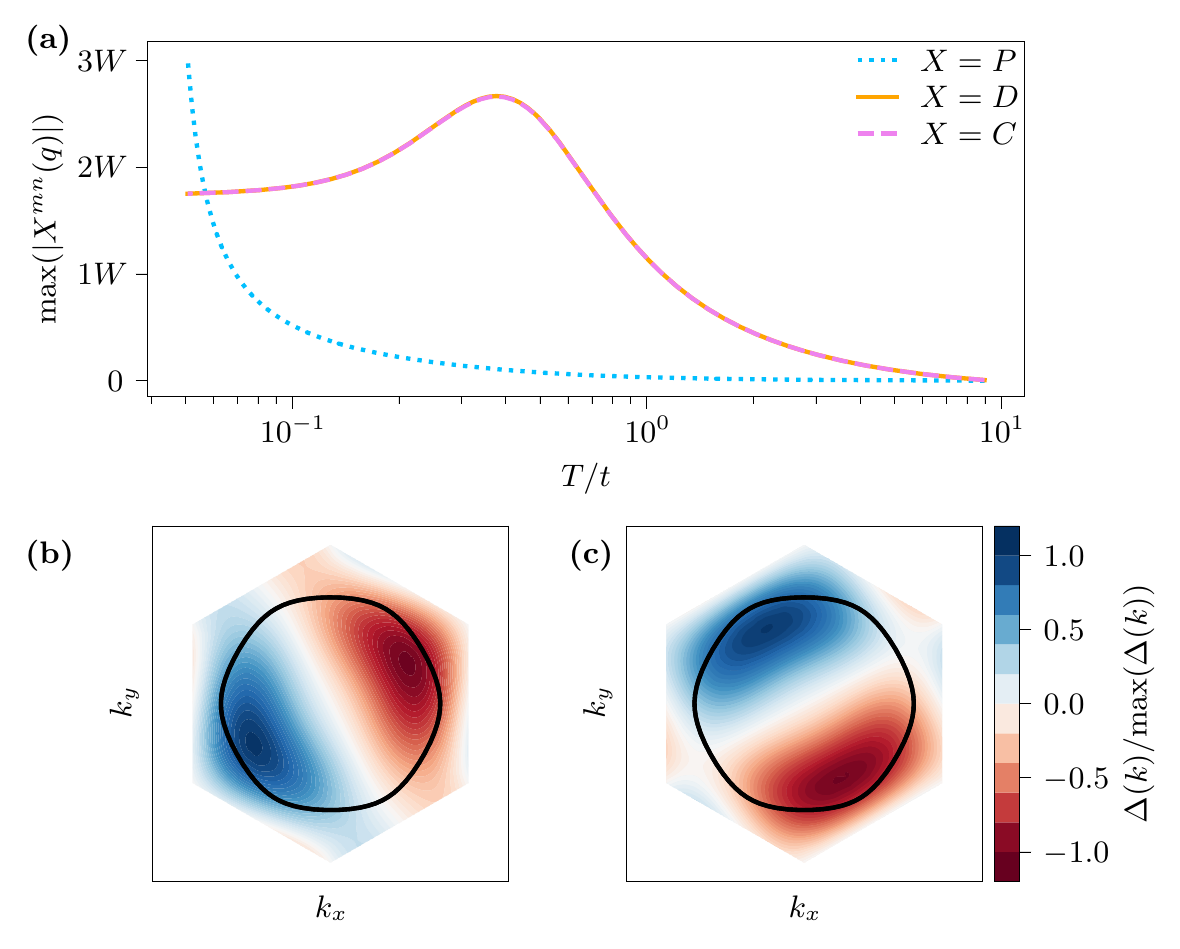}
    \caption{\textbf{TUFRG results for $V_1 / t = -1$ and $\mu / t = 1.2$ with $N_{\bm{q}}=540$, $N_f=19$} For even lower fillings we still find a divergence of the $P$ channel (a). But the reconstructed (degenerate) leading gaps $\Delta(k)$ of the emerging superconductivity instability (see (b)) depict now functions in the $E_1$ irrep. of $C_{6v}$. The black line represents the Fermi surface, featuring 2 zero crossings.}
    \label{fig:attractivePanel12}
\end{figure}

Lowering $\mu$ further, $p$-wave superconductivity becomes the leading instability, which is described by the two-dimensional $E_1$ irrep. of the same point group (see Figs.~\ref{fig:attractive12_patching} and \ref{fig:attractivePanel12}). 
On a mean-field level, it is energetically beneficial for the superconducting order to open a full gap in the quasi-particle spectrum, which can be accomplished, for example, by constructing the superconducting gap $\Delta(\bm{k})$ as a complex superposition of the $E_1$ lattice harmonics. 
This leads to a  $p+ip$ superconducting state featuring a finite Chern number $\mathcal{C} = -1$ which is thus topologically non-trivial (see App.~\ref{app:chern}).

Qualitatively, the two superconducting instabilities we find here are also consistent with the mean-field study presented in Ref.~\cite{PhysRevB.81.024504}. We note, however, that our FRG study includes additional fluctuations, which induce the Pomeranchuk instability when approaching Van Hove filling.

\subsection{Phase diagram of the attractive case}

\begin{figure}[t!]
    \includegraphics[width=\columnwidth]{ 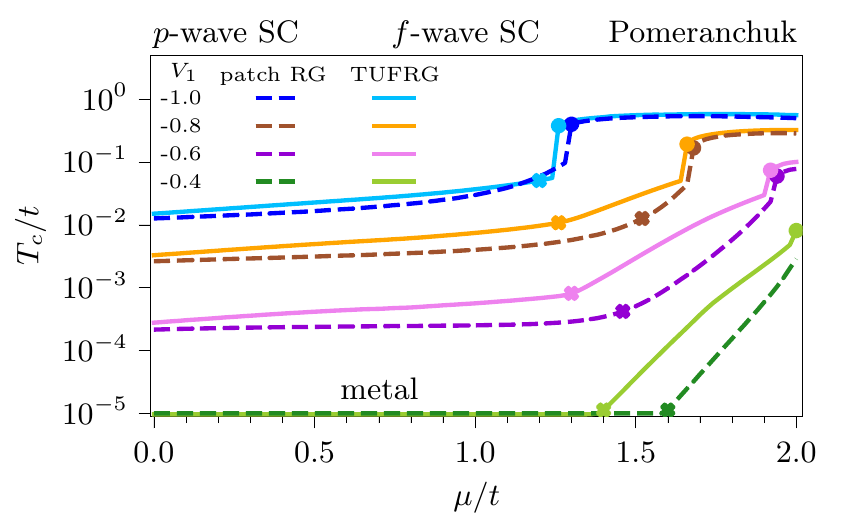}
    \caption{\textbf{Phase diagram for attractive interactions from patching and TUFRG.} At Van Hove filling and for sufficiently strong interactions, both methods consistently predict a Pomeranchuk instability (see Figs.~\ref{fig:attractive20_patching} \& \ref{fig:attractivePanel20} for more details). Below $\mu / t = 2.0$, two kinds of pairing instabilities can be found: an $f$-wave superconductor in vicinity of Van Hove filling (see Figs.~\ref{fig:attractive18_patching} \& \ref{fig:attractivePanel18}) and a $p$-wave instability (see Figs.~\ref{fig:attractive12_patching} \& \ref{fig:attractivePanel12}) at even smaller values of $\mu / t$. The boundaries are indicated by colored crosses (for the $f$-wave superconductor) or dots (for the Pomeranchuk instability), respectively.}
    \label{fig:attractive}
\end{figure}

In Fig.~\ref{fig:attractive}, we have mapped out the phase diagram for various $V_1 / t < 0$ using both, the $N$-patch FRG and the TUFRG. Generally, the phase boundaries, the respective ground state instabilities, and the critical scales are in reasonable agreement.
Some deviations in the critical temperatures are visible, in particular, in the regions where the superconducting instabilities occur at very low scales. 
Notably, the transition from Pomeranchuk to the \textit{f}-wave superconductivity is in good alignment in both methods while the second transition point towards \textit{p}-wave superconductivity has a larger difference although deep into this particular phase the methods apparently converge.

To establish the reliability of our results, we have further studied the convergence of the TUFRG approach with respect to the momentum- and form-factor resolution $(N_{\bm{q}},N_f)$ in more detail, see App.~\ref{app:momentaff}.

\section{Repulsive case $V_1 > 0$}

We now consider the repulsive case $V_1 / t > 0$.
Here, we can expect 
that the occurring instabilities result from an interplay of the perfect nesting at the Van Hove point, whose effect can be mitigated by changing the filling, and a divergent susceptibility in the pairing channel, which eventually induces a superconducting instability. 

\subsection{CDW at Van Hove filling}

\begin{figure}[t!]
    \includegraphics[width=\columnwidth]{ 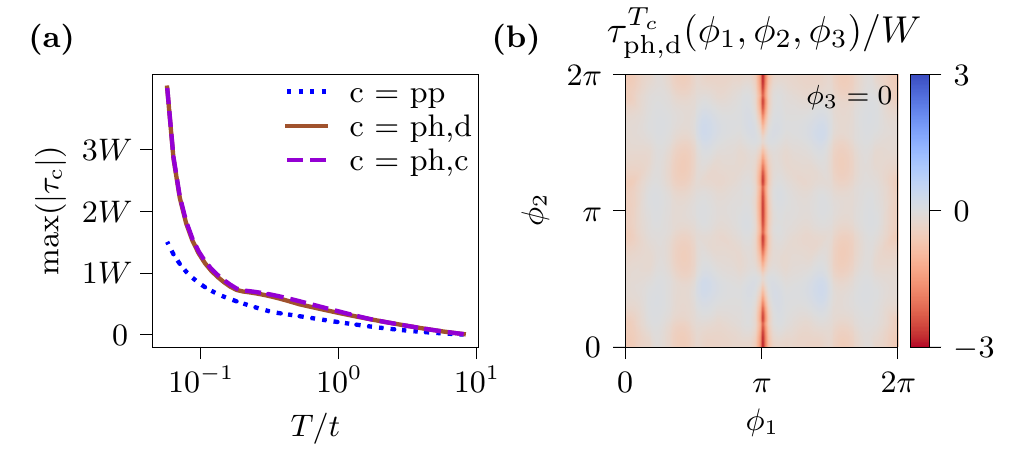}
    \caption{\textbf{Patching results for $V_1 / t = 1$ at Van Hove filling}. (a) Flow of the channel maxima, indicating a simultaneous divergence in both particle-hole channels, consistent with the TUFRG result in Fig.~\ref{fig:repulsivepanel20}. (b) Plot of the direct particle-hole channel right at the critical scale $T_c$. The corresponding plot for $\tau_{\mathrm{ph,c}}$ can be obtained via crossing symmetry, i.e. a permutation $\phi_1$ and $\phi_2$ and a flip of the overall sign.}
    \label{fig:repulsive20_patching}
\end{figure}

\begin{figure}[t!]
    \includegraphics[width=\columnwidth]{ 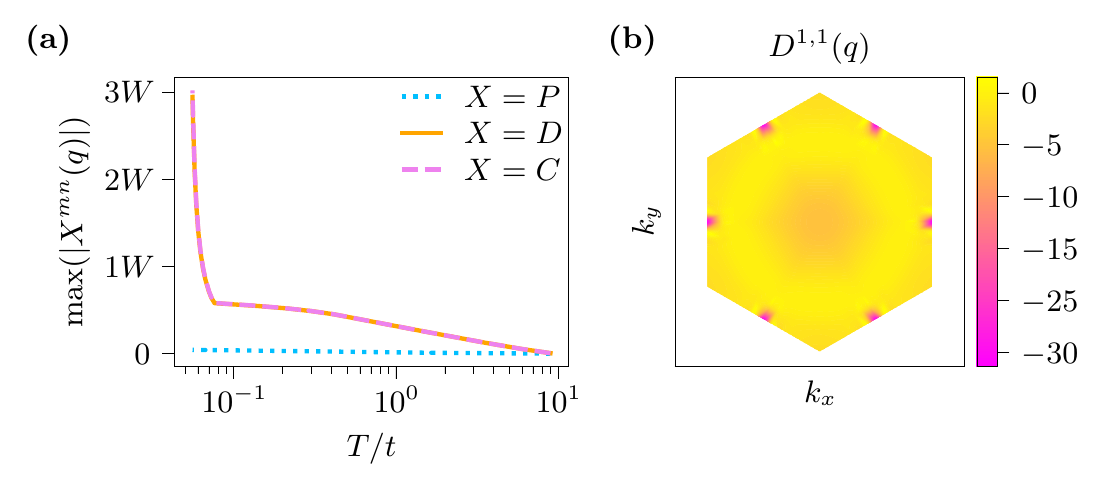}
    \caption{\textbf{TUFRG results for $V_1 / t = 1$ and $\mu / t = 2$ with $N_{\bm{q}}=540$, $N_f=19$.} Tracking the evolution of the channels $P,C,D$, we will find a CDW instability as the maximal absolute value of the $C$ and $D$ diverge while the $P$ channel remains small, see (a). The alignment of the $C$ and $D$ channel is still expected because of the symmetric connection of the diagrams.  The on-site, momentum resolved $D$ channel $D^{1,1}(q)$ inhabits peaks at the $\bm{M}$ points, indicating the emergence of the CDW with modulation $\exp{(i\bm{M}\bm{R}}$).}
    \label{fig:repulsivepanel20}
\end{figure}

Similar to the attractive case, both 
methods detect a divergence of the particle-hole channels for $\mu /t = 2$. An analysis of the possible order parameters $\langle \bpsi_{\bm{k} + \bm{q}} \psi_{\bm{k}} \rangle$ (see Figs.~\ref{fig:repulsive20_patching} and~\ref{fig:repulsivepanel20}), however, reveals that the leading instability occurs for transfer momenta $\bm{q}$, which coincide with the nesting vector $\bm{M}$. The FRG results thus 
indicates the instability towards a charge density wave.

\subsection{$\tilde{p}$-wave superconductivity below Van Hove filling}

As we have discussed for the attractive case, the Fermi surface loses its nesting property below Van Hove filling, and, thus, fluctuations in the particle-hole channels are weaker (but still finite). In contrast to our previous considerations, however, putative superconducting instabilities would now arise from a different mechanism. Since $V_1 / t > 0$, pairing is not directly encapsulated by the bare vertex and an attractive interaction in $\tau_{\mathrm{pp}}$ henceforth needs to be generated by inter-channel feedback during the RG flow. 

\begin{figure}[t!]
    \includegraphics[width=\columnwidth]{ 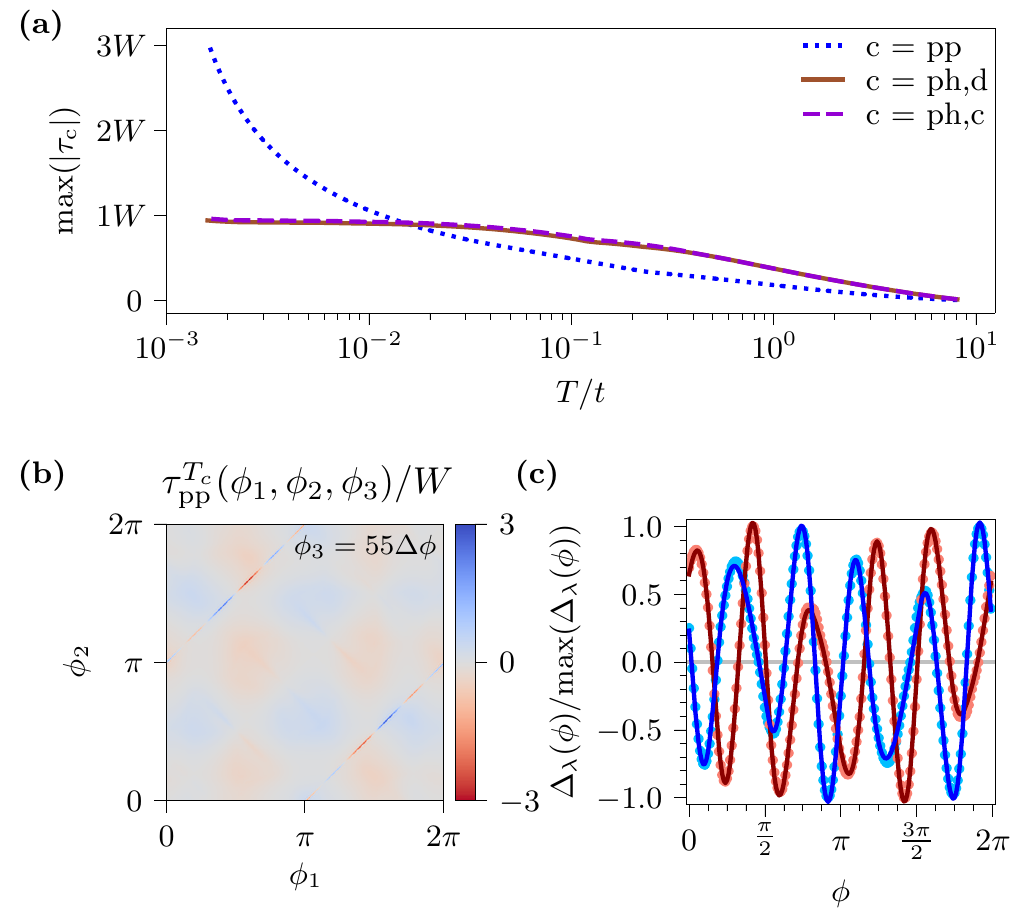}
    \caption{\textbf{Patching results for $V_1 / t = 1$ and $\mu / t = 1.7$.} A superconducting instability, driven by strong particle-hole fluctuations, becomes visible as a divergence of the particle-particle channel (see (a) and (b)). The pairing potential, constructed from $\tau_{\mathrm{pp}}$ at the critical scale $T_c$, has two degenerate gaps (light red and light blue dots in (c)), which can be fit by a linear combination of first and second neighbor lattice harmonics of the two-dimensional $E_1$ representation of $C_{6v}$ (dark red/blue line).}
    \label{fig:repulsive17_patching}
\end{figure}

\begin{figure}[t!]
    \includegraphics[width=\columnwidth]{ 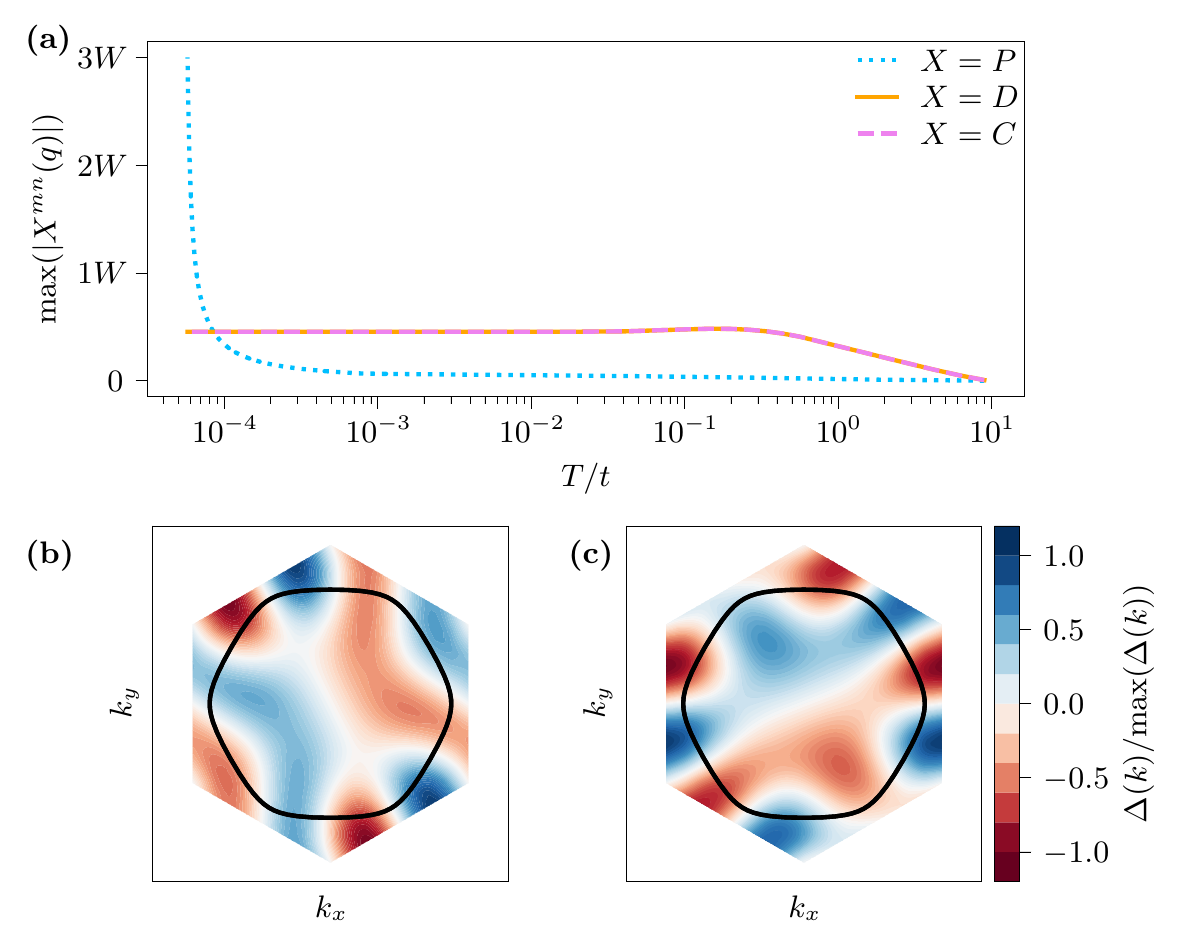}
    \caption{\textbf{TUFRG results for $V_1 / t = 1$ and $\mu / t = 1.7$ with $N_{\bm{q}}=540$, $N_f=19$} The RG flow for repulsive interactions away from Van Hove filling features the divergence of the $P$ channel and hence a superconductive instability (a).  The reconstructed degenerate leading gaps $\Delta(k)$ of this instability (see (b)) depict a higher harmonic function of the $E_1$ irrep. of $C_{6v}$. The black line represents the Fermi surface, featuring 10 zero crossings each.}
    \label{fig:repulsivepanel17}
\end{figure}

Indeed, both methods find an instability of the particle-particle channel for various fillings $\mu / t < 0$ and, remarkably, the flows of the maxima in the different channels plotted in Figs.~\ref{fig:repulsive17_patching}(a) and~\ref{fig:repulsivepanel17}(a) underline the importance of particle-hole fluctuations for the emergence of superconductivity. While the pairing channel is negligible (in TUFRG) or at least smaller than the other contributions (in the patching scheme), the particle-hole channels first sharply increase and then converge to a constant value, which dominates the vertex. In the low temperature regime, however, an abrupt upturn in the $\tau_{\mathrm{pp}}$ flow can be observed, which ultimately results in a divergence of the RG flow. The respective gap function again transforms in the $E_1$ representation of $C_{6v}$, but requires both nearest- and second-nearest neighbor lattice harmonics, as indicated by an increased number of nodes on the Fermi surface (see Fig.~\ref{fig:repulsive17_patching}(c) or  Fig.~\ref{fig:repulsivepanel17}(b)). We dub this instability $\tilde{p}$-wave to set it apart from its counterpart in the attractive case.

Notably, a complex order parameter constructed solely from the second neighbor $E_1$ basis functions likewise yields $\mathcal{C} = -1$, whereas superpositions of both the first and second neighbor harmonics can generate an enhanced quantum Hall response due to Chern numbers $|\mathcal{C}| > 1$ (see Fig.~\ref{fig:Chern} for more details). 

\subsection{Phase diagram of the repulsive case}

\begin{figure}[t!]
    \includegraphics[width=\columnwidth]{ 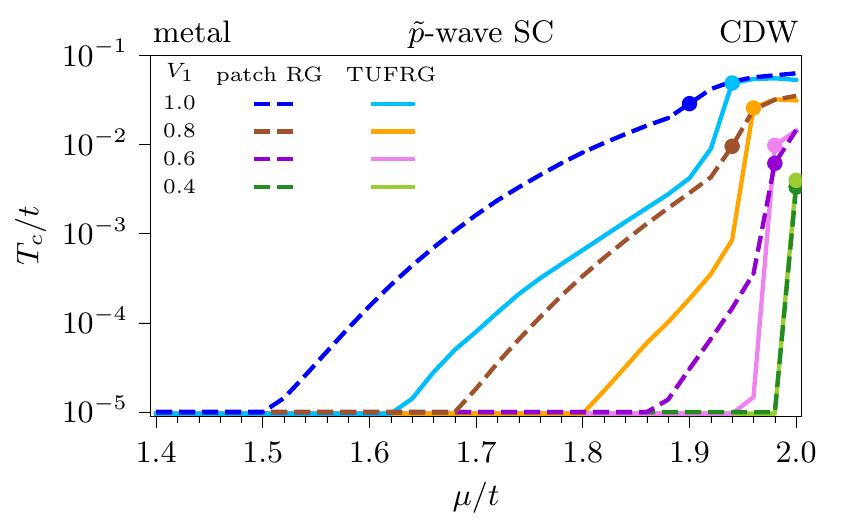}
    \caption{\textbf{Phase diagram for repulsive interactions from patching and TUFRG.} Both approaches predict one transition from a metallic state, where no instability of the RG flow is observed down to $T / t = 10^{-5}$, to an extended $\tilde{p}$-wave superconductor (see Figs.~\ref{fig:repulsive17_patching} \& \ref{fig:repulsivepanel17}), followed by another transition (indicated by a colored dot) to a charge density wave with transfer momentum $\bm{q} = \bm{M}$ close to Van Hove filling (see Figs.~\ref{fig:repulsive20_patching} \& \ref{fig:repulsivepanel20}).}
    \label{fig:repulsivePhasediagram}
\end{figure}

In Fig.~\ref{fig:repulsivePhasediagram} we finally show results for the phase diagram obtained from the patching scheme and TUFRG for various fillings and repulsive interactions. Interestingly, the temperature scales for the $\tilde{p}$-wave superconductor measured in the patching scheme are almost one order of magnitude higher than in TUFRG, though the nature of the instability remains the same. Moreover, the sharp drop in $T_c$ between the CDW and superconducting regime is absent in the patching results, where only a soft shoulder is indicative of the transition. Close to Van Hove filling on the other hand, the agreement is more reasonable. Since the central patch points coincide with the saddle points in the latter case, this generates the suspicion that the projection to the Fermi surface might be responsible for the observed discrepancy away from perfect nesting.




\section{Discussion}
We analyzed competing orders in a model of spinless electrons on the triangular lattice with nearest-neighbor interaction.
Our study was motivated by the observation of correlated states in moir\'e bilayers of transition metal dichalcogenides. 
These systems are effectively described by interacting electrons on a triangular lattice, although equipped with (pseudo)spin and/or orbital degrees of freedom. 
To distill out the minimal degrees of freedom, we considered the paradigmatic toy model of spinless electrons and showed that it still possesses a rich interplay of ordering tendencies in the vicinity of a Van Hove singularity. To resolve this interplay, we calculated the effective two-particle interaction vertex in an unbiased way with the functional renormalization group. It is crucial to accurately resolve the momentum dependence of the vertex and we used two different parameterizations - a patching scheme for the Fermi surface and a channel decomposition for the momentum transfers. Both of them give qualitatively consistent results. 

With an attractive bare interaction, we find a Pomeranchuk instability in the $s$-wave channel directly around Van Hove filling and $f$- and $p$-wave pairing instabilities in its vicinity for smaller fillings. Within RPA, both the charge and pairing channel can develop an instability, although at weak coupling the pairing channel has a stronger divergence (logarithmic vs double logarithmic). Interestingly, in our calculations, the Pomeranchuk instability in the charge channel develops first due to non-universal effects (beyond the logarithmic scaling). 
The $s$-wave Pomeranchuk instability corresponds to a singular compressibility but is not associated with any symmetry-breaking order. This can signal the tendency to phase separation with domains of different density. Another possibility is that the divergence is cured by terms outside of our truncation, e.g., by self-energy terms, and makes room for a subleading instability. 
The p-wave pairing solution is two-fold degenerate and can form chiral $p+ip$ superconductivity in the ground state. This topological triplet superconducting state breaks time-reversal symmetry and can host Majorana modes on its boundaries. 

In the case of a repulsive bare interaction, we obtain a CDW instability closest to Van Hove filling, whose fluctuations mediate unconventional $p$-wave pairing at smaller fillings. The wave vectors of the CDW are the three non-equivalent $M$ points of the Brillouin zone and the exact charge pattern of the associated order depends on their combination in the ground state. Due to the bare nearest-neighbor repulsion, we find the unconventional p-wave pairing to be of extended size described by nearest- and next-nearest-neighbor harmonics. This can yield topological $p+ip$ states with higher Chern numbers, which increases, e.g., the number of chiral edge modes and the quantum Hall response.  


\begin{acknowledgement}
\textit{Acknowledgements}.--- We thank P. Bonetti for helpful discussions. R.T. is funded by  Deutsche
Forschungsgemeinschaft (DFG, German Research Foundation) through
Project-ID 258499086 - SFB 1170 and through the W\"urzburg-Dresden
Cluster of Excellence on Complexity and Topology in Quantum Matter - ct.qmat Project-ID 390858490 - EXC 2147.
N.G., D.K., and M.M.S. acknowledge support through the SFB 1238 (project C02, project id 277146847).
M.M.S. is funded by the DFG Heisenberg programme (project id 452976698).

\end{acknowledgement}

\section*{Author contributions}
NG implemented the TUFRG code and ran the respective calculations. DK implemented the patch code and ran these calculations. RH ran additional TUFRG calculations and helped in the benchmarking process. RT, LC and MMS supervised  the study. All authors were involved in the preparation of the manuscript.

\subsection*{Data availability statement }
This manuscript has no
associated data or the data will not be deposited. [Authors’
comment: The data generated in this study for both methods is available from the respective authors on reasonable request.]

\bibliography{spinless}
\clearpage


\onecolumn
\appendix

\section{Temperature regulator}
\label{app:regulator}

Both codes applied in this manuscript make use of the temperature flow scheme developed by Honerkamp and Salmhofer \cite{Honerkamp_Tflow}. As such, the bare propagator is regularized as 
\begin{align}
    G_0(i\omega,\bm{k}) \rightarrow  G_0^T(i\omega,\bm{k})= \frac{T^{1/2}}{i\omega-\xi(\bm{k})} \,,
\end{align}
while the fermionic fields $\bpsi, \psi$ are simultaneously rescaled by a factor $T^{-3/4}$. This way, the temperature only appears in the Gaussian part of the action and the flow equations \eqref{eq:flowequation} apply up to a substitution $\Lambda \to T$. Another prominent advantage of this regularization, apart from being able to directly identify $\Lambda$ with a physical quantity (temperature), is that contributions from particle-hole loops are fully taken into account even for small total momenta. In contrast to, for example, momentum shell schemes (see Ref.~\cite{RevModPhys.84.299}), instabilities with transfer momenta at the $\bm{\Gamma}$ point are therefore not artificially suppressed, allowing to treat all channels in an unbiased way \cite{Honerkamp_Tflow}.

\section{Flow equations and numerical implementation of TUFRG}\label{app:flow_eq_TUFRG}
\subsection{Elements of flow equations}\label{app:details}
Each flow equation Eqs.~\eqref{eq::ffflowP}-\eqref{eq::ffflowD} consists of a product of  a particle-particle ($-$) or particle-hole ($+$) bubble integral $\dot{B}(\bm{q})^{\pm}_{l,l'}$ connecting two cross-channel projections $V^{X}$, with $X=P,C,D$. For completeness, both objects will be described here explicitly. 
The bubble integrals emerge by insertion of the form-factor resolved unities in Eqs.~\eqref{eq::contribution1}-\eqref{eq::contribution3} to separate the loops of the diagrams from the vertices. Their explicit form is given by
\begin{align}
        \dot{B}(\bm{q})_{l,l'}^{(\pm)} =- \int_{p}\frac{d}{d\Lambda}  [G_0^{\Lambda}( i\omega,\bm{q}+\bm{p})\times G_0^{\Lambda}(\pm i\omega,\pm \bm{q}) ] f_l(\bm{p})f^{*}_{l'}(\bm{p})\,. 
        \label{eq::bubbleform}
\end{align}
By implementing the temperature flow as shown in App.~\ref{app:regulator} and performing the Matsubara summations explicitly the bubbles are cast into:
\begin{align}
 \dot{B}(\bm{q})_{l,l'}^{(+)}\! &=\!+\!\!\int_{\bm{p}}\! \frac{n_F'(\xi(\bm{q}\!+\!\bm{p})\!)\!-\!n_F'(\xi(\bm{q})\!) }{ \xi(\bm{q}\!+\!\bm{p})\!-\! \xi(\bm{q})}f_l(\bm{p})f_{l'}^{*}(\bm{p}),         \label{eq::bubble1} \\[8pt]
 \dot{B}(\bm{q})_{l,l'}^{(-)}\! &=\!-\!\!\int_{\bm{p}}\!\! \frac{n_F'(\xi(\!\bm{q}\!+\!\bm{p})\!)\!+\!n_F'(\xi(\!-\bm{q})\!)}{ \xi(\bm{q}\!+\!\bm{p})\!+\! \xi(-\bm{q})}f_l(\bm{p})f_{l'}^{*}(\bm{p}),         \label{eq::bubble2}
\end{align}
where $n_F'(x)$ is the Fermi function after performing the temperature-derivative i.e. $n_F'(x) = \frac{d}{dT} n_F(x)$.
After inserting the form-factor resolved unities into the initial flow equations, the vertices will also gain a dependency on the form-factors. The emergent objects will be the cross-channel projections:
\begin{align}
V^{P}_{l,l'}(\bm{q})&\!=\!\! \int_{\bm{k},\bm{k}'}\!\! f_l(\bm{k})f_{l'}^{*}(\bm{k}') V^{\Lambda}(\bm{k}\!+\! \bm{q},\!-\bm{k}, \bm{k}'\!+\!\bm{q},\!-\bm{k}'),\label{eq::crossprojection0}\\
V^{C}_{l,l'}(\bm{q})&\!=\!\!\int_{\bm{k},\bm{k}'}\!\!  f_l(\bm{k})f_{l'}^{*}(\bm{k}') V^{\Lambda}(\bm{k} + \bm{q}, \bm{k}', \bm{k}' + \bm{q},\bm{k})\,,\\
V^{D}_{l,l'}(\bm{q})&\!=\!\! \int_{\bm{k},\bm{k}'}\!\!  f_l(\bm{k})f_{l'}^{*}(\bm{k}') V^{\Lambda}(\bm{k} + \bm{q}, \bm{k}', \bm{k} ,\bm{k}'+\bm{q})\,,                \label{eq::crossprojection}
\end{align}
where the integral includes the Brillouin zone area: $\int_{\bm{k}}=A_{\mathrm{BZ}}^{-1}\int d\bm{k}$. These expressions can also be simplified by plugging in the plane wave form-factors $\exp(i\bm{k}\bm{R}_l)$ (see App.~\ref{app:momentaff}) and expressing $V^{\Lambda}$ by the decomposition Eq.~\eqref{eq::decomposition}. Therefore the double integral over the Brillouin zone is exchanged by a simple sum over the selected form-factors $\sum_{L}$:
\begin{align}\label{eq::crossprojectionsexplicit1}
V^{P}_{l,l'}(\bm{q})\! =\! V^{P,0}_{l,l'}(\bm{q})\!+\!V^{P\veryshortarrow C}_{l,l'}(\bm{q})\!+\!V^{P\veryshortarrow D}_{l,l'}(\bm{q})\!+\!P_{l,l'}(\bm{q})\,,
\end{align}
\begin{align}
V^{P\veryshortarrow C}_{l,l'}(\bm{q}) &=\! \sum_{L} \!\tilde{C}_{\bm{R}_{L},-\bm{R}_{L}+\bm{R}_{l}+\bm{R}_{l'}} (\!-\!\bm{R}_{L}\!+\!\bm{R}_{l'}) e^{-i(\bm{R}_{L}\!-\!\bm{R}_{l'})\bm{q}}\,,         \nonumber \\ 
V^{P\veryshortarrow D}_{l,l'}(\bm{q}) &=\! \sum_{L}\! \tilde{D}_{\bm{R}_{L},-\bm{R}_{L}+\bm{R}_{l}-\bm{R}_{l'}} (\!-\!\bm{R}_{L}\!-\!\bm{R}_{l'}) e^{-i\bm{R}_{L}\bm{q}}\,.\nonumber
\end{align}
\begin{align}
V^{C}_{l,l'}(\bm{q})\!=\!V^{C,0}_{l,l'}(\bm{q})\!+\!V^{C \veryshortarrow P }_{l,l'}(\bm{q})\!+\!V^{C \veryshortarrow D}_{l,l'}(\bm{q})\!+\! C_{l,l'}(\bm{q})\,,
\end{align}
\begin{align}
V^{C\veryshortarrow P}_{l,l'}(\bm{q}) &=\!  \sum_{L}   \tilde{P}_{\bm{R}_{L},-\bm{R}_{L}+\bm{R}_{l}+\bm{R}_{l'}} (\!-\!\bm{R}_{L}\!+\!\bm{R}_{l'}) e^{-i(\bm{R}_{L}\!-\!\bm{R}_{l'})\bm{q}}\,,\nonumber\\
V^{C\veryshortarrow D}_{l,l'}(\bm{q}) &=\! \sum_{L}    \tilde{D}_{\bm{R}_{L},\bm{R}_{L}-\bm{R}_{l}+\bm{R}_{l'}} (-\bm{R}_{l})e^{-i\bm{R}_{L}\bm{q}}\,,\nonumber
\end{align}
\begin{align}\label{eq::crossprojectionsexplicit3}
V^{D}_{l,l'}(\bm{q})\!=\!V^{D,0}_{l,l'}(\bm{q})\!+\! V^{D \veryshortarrow P}_{l,l'}(\bm{q})\!+ \!V^{D \veryshortarrow C}_{l,l'}(\bm{q})\!+\! D_{l,l'}(\bm{q})\,,
\end{align}
\begin{align}
V^{D\veryshortarrow P}_{l,l'}(\bm{q}) &= \sum_{L} \tilde{P}_{\bm{R}_{L},\bm{R}_{L}-\bm{R}_{l}-\bm{R}_{l'}} (-\bm{R}_{l}) e^{-i(\bm{R}_{L} - \bm{R}_{l'})\bm{q}}\,,  \nonumber \\
V^{D\veryshortarrow C}_{l,l'}(\bm{q}) &= \sum_{L} \tilde{C}_{\bm{R}_{L},\bm{R}_{L}-\bm{R}_{l}+\bm{R}_{l'}} (-\bm{R}_{l}) e^{-i\bm{R}_{L}\bm{q}}.
\nonumber
\end{align}
The objects $V^{X,0}_{l,l'}(\bm{q})$ encode the initial interaction of the model Eq.~\eqref{eq:model} by projecting it into the respective channels, see App.~\ref{app:initial}. 
$\tilde{X}_{l,l'}$ represents the Fourier-transformed channels, for example for the pairing channel $P$:
\begin{align}
    \tilde{P}_{l,l'}(R_i) = A^{-1}_{\mathrm{BZ}} \int \ d\bm{p}\, P_{l,l'}(\bm{p}) e^{-i\bm{p}\bm{R}_i} .
\end{align}

\subsection{Choice of momenta and form-factors and convergence}\label{app:momentaff}
\begin{figure}[t!]
    \centering
    \includegraphics[width=0.8\columnwidth]{ 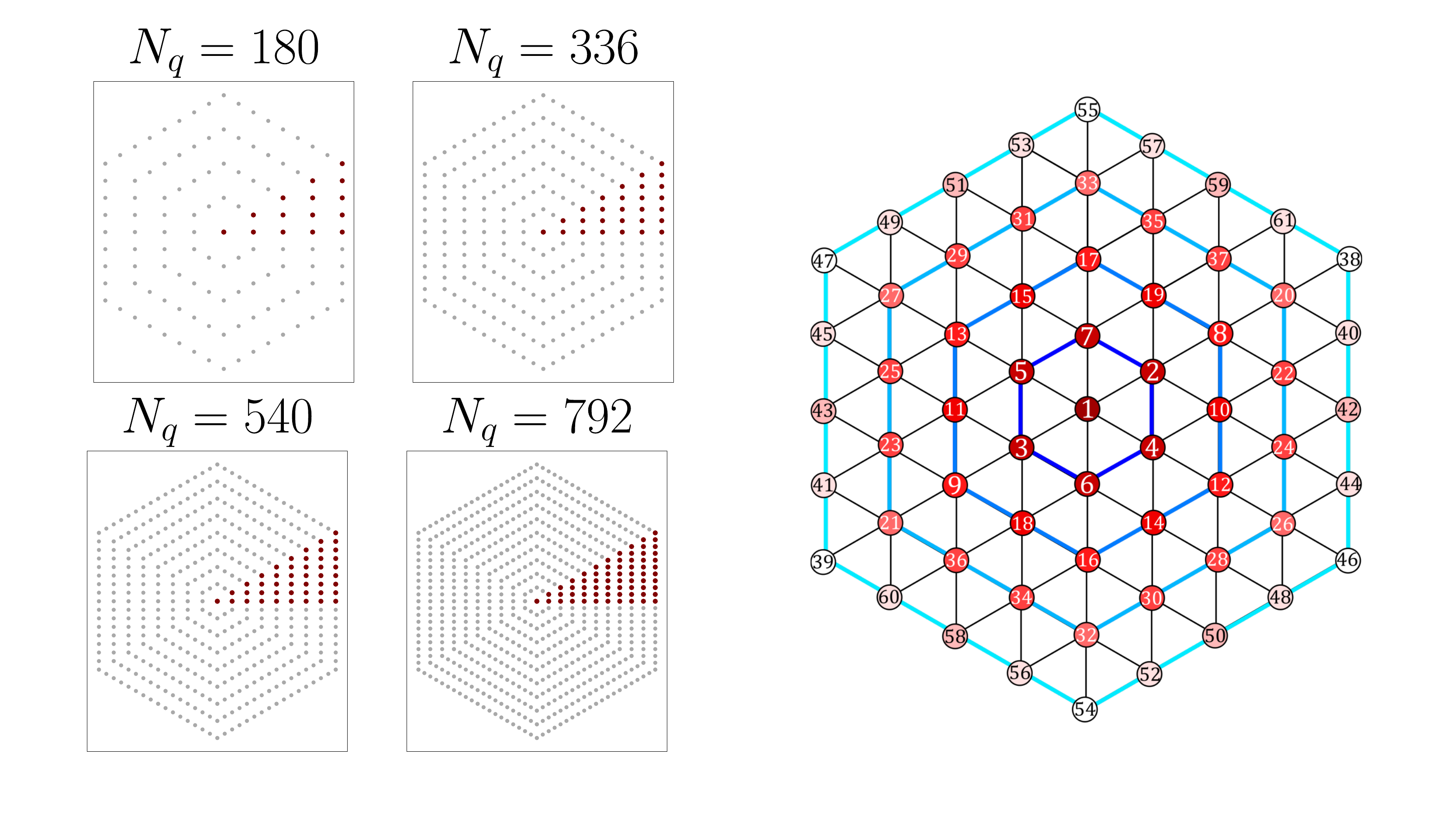}
    \caption{\textbf{Momentum resolutions and form-factors choice.} Left: different resolutions of the Brillouin zone. Only $1/12$ of the momenta (red) actually have to be calculated in the RG flow while the rest can be derived by symmetry relations. Right: real space vectors $\bm{R}_l$ for the plane wave form-factors. }
    \label{fig:momentaff}
\end{figure}

\begin{figure}[t!]
    \centering
    \includegraphics[width=0.8\columnwidth]{ 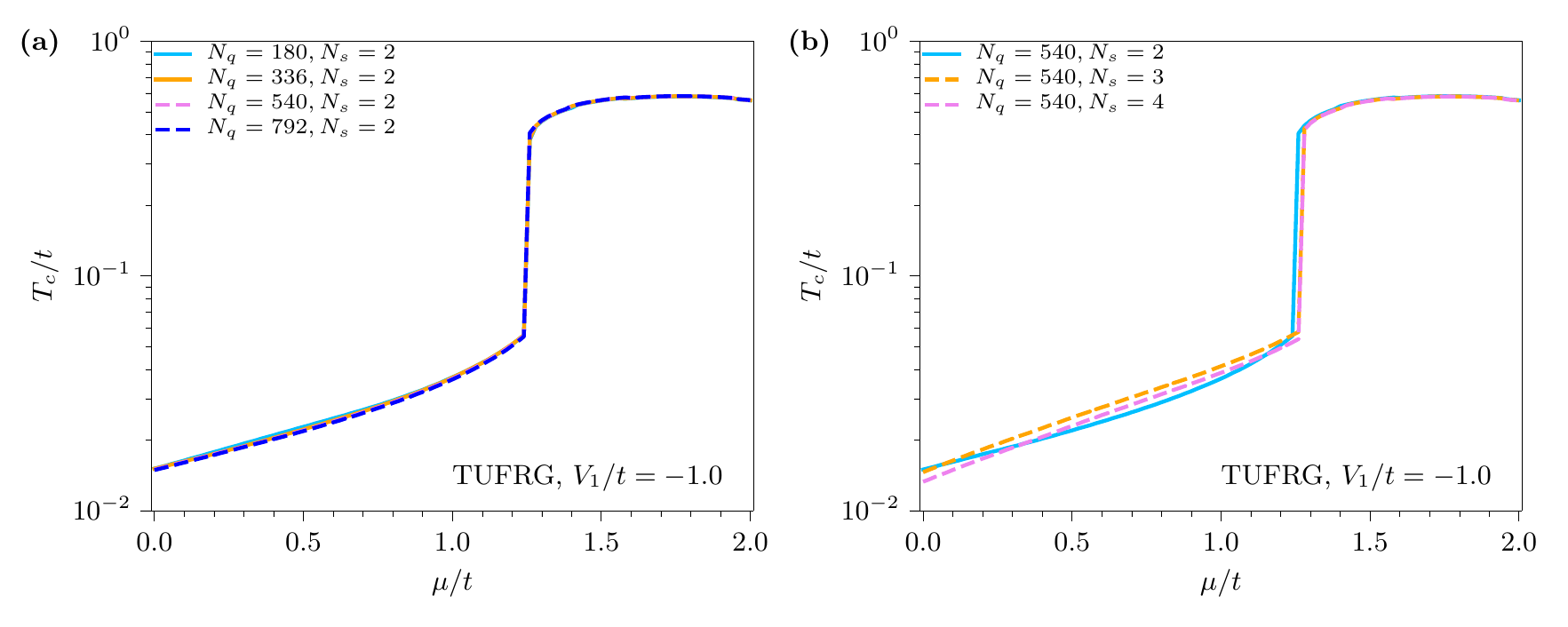}
    \caption{\textbf{Convergence of critical RG scales from TUFRG for the attractive case $V_1 / t < 0$.} (a) Study for convergence for increasing momentum resolution. All calculations align qualitatively and quantitatively for the checked region. (b) Study for convergence in form factors. While the results match qualitatively, minor deviations in the critical temperature regarding the superconductive instabilities occur. }
    \label{fig:attractiveConvergence}
\end{figure}

\begin{figure}[t!]
    \centering
    \includegraphics[width=0.8\columnwidth]{ 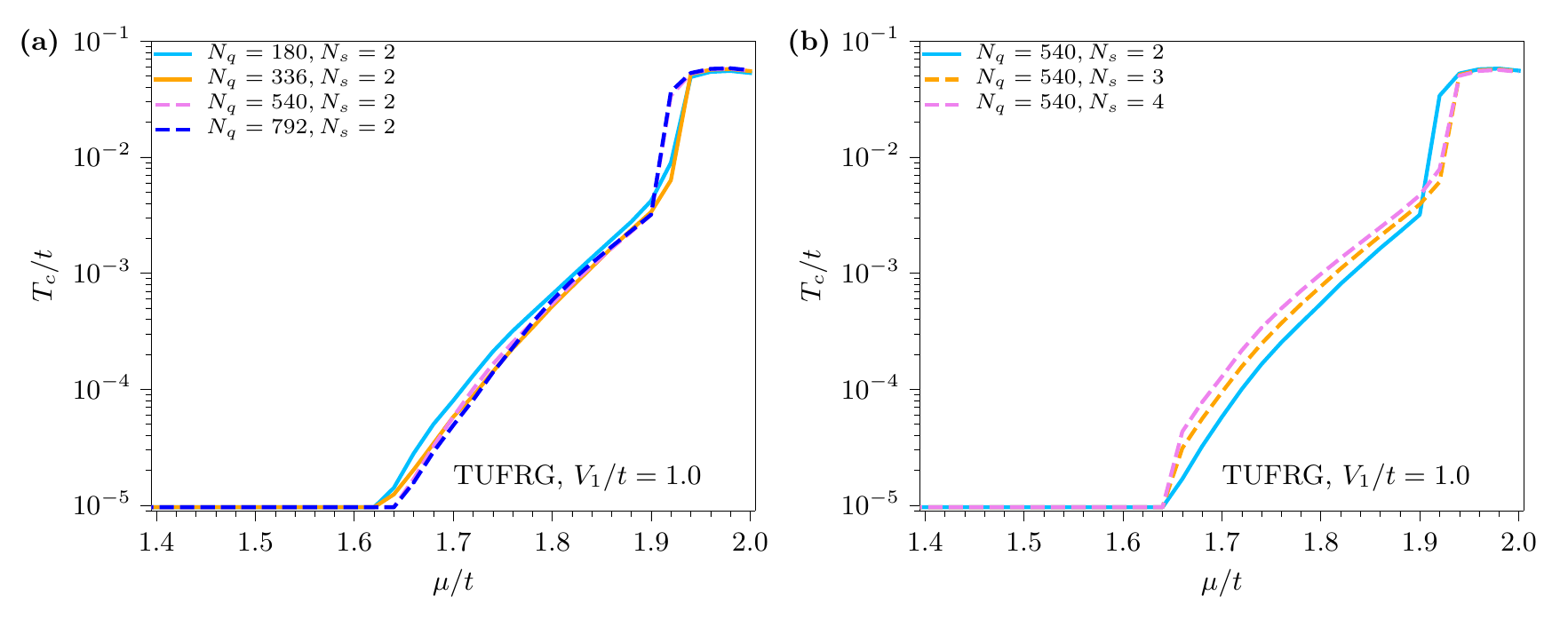}
    \caption{\textbf{Convergence of critical RG scales from TUFRG for the repulsive case $V_1 / t >0$.}(a) For the investigation of convergence in increasing momentum resolution $N_{\bm{q}}$ we find qualitatively the same phase diagram which quantitative deviations diminish for higher resolution.  (b) The the investigation of including more form factors $N_{f}$ we still find qualitative alignment, while the critical temperature slightly grows for including more shells. Since we are primarily interested in the qualitative behaviour and the deviations are not too strong, we use $N_s=2$ for the calculations in the sections 4 and 5. }
    \label{fig:repulsiveConvergence}
\end{figure}

One has the freedom to select different sets of form-factors as long as the  unity condition Eqs.~\eqref{eq::unity1}-\eqref{eq::unity}  are fulfilled. The simplest choice of form-factors have the form of plane waves: $f_l(\bm{k})=\exp\left(i\bm{k}\bm{R}_l \right)$ where $\bm{R}_l$ is a real space vector of the lattice of the investigated model, i.e. in our case the triangular lattice. This choice has the advantage, that the truncation of form-factors can be done within an interpretable reasoning: the inclusion of a form-factor $f_l(\bm{k})$ will correspond to taking effects of fermionic bilinears with distance $\bm{R}_l$ into account \cite{PhysRevB.98.155132}. Since we assume that the emerging physics in the RG flow will be predominately influenced by short-range effects, we will truncate all form-factors which exceed a chosen distance. In our calculations we mostly select $N_f=19$ form-factors, corresponding to on-site (i.e. $\bm{R}_1=0$), first-, second- and third-nearest neighbors effects. For the convergence checks in Figs.~\ref{fig:attractiveConvergence},\ref{fig:repulsiveConvergence} we will also use $N_f=37$ (i.e. up to 5th nearest-neighbors effects) and $N_f=61$ (i.e. up to 8th nearest-neighbors effects), see Fig.\ref{fig:momentaff}. This specific choice of amount of form-factors is based on keeping a \textit{hexagonal-shell} $N_s$ into account. This means, that we will include all plane waves with $\bm{R}_l$  which are on or inside the $N_s-th$ hexagon of the real space lattice, cf. Fig.\ref{fig:momentaff}. Therefore the numbers $N_f=19,37,61$ correspond to the hexagon-shells $N_s=2,3,4$.

For the momentum resolution, we choose evenly placed points in the Brillouin zone. Most of our calculations are done with $N_{\bm{q}}=180$ momenta to compare it with the $192$ patching points of the other approach, while for the convergence checks in Figs.~\ref{fig:attractiveConvergence},\ref{fig:repulsiveConvergence} we also choose $N_{\bm{q}}=336$, $N_{\bm{q}}=540$ and $N_{\bm{q}}=792$.

Actually, one does not have to calculate the RG flow for all momenta $N_{\bm{q}}$, but only for a fraction $1/12 \times N_{\bm{q}}$. The rest of the contributions can then be restored by symmetry relations since the symmetries of the initial model Eq.~\eqref{eq:model} are inherited by the flow equations, see \cite{PhysRevB.103.235150} for details.


\section{Initial conditions}
\label{app:initial}

The initial condition for the FRG flow is given by the bare two-particle vertex $V_0$, which can be directly read off the microscopic model in Eq.~\eqref{eq:model}. For this purpose, one needs to identify the action $S_{\text{int}}$ with the vertex at the UV scale, i.e $S_{\text{int}}=V^{\Lambda_{\text{UV}}} = V_0$, and additionally account for crossing symmetries, such as $V(\bm{k}_1, \bm{k}_2, \bm{k}_3, \bm{k_4}) = - V(\bm{k}_2, \bm{k}_1, \bm{k}_3, \bm{k_4})$. The initial condition needs to be properly (anti-) symmetrized henceforth. On the level of the Hamiltonian, crossing symmetry can already be made explicit by reordering the Fock space operators as
\begin{equation}
   V_1 \underset{\langle i j\rangle}{\sum}n_i n_j=V_1 \underset{\langle i j\rangle}{\sum}c_i^\dagger c_j^\dagger c_j c_i=\frac{1}{4} V_1\underset{\langle i j\rangle}{\sum}\Bigg(c_i^\dagger c_j^\dagger c_j c_i-c_i^\dagger c_j^\dagger c_i c_j-c_j^\dagger c_i^\dagger c_j c_i+c_j^\dagger c_i^\dagger c_i c_j\Bigg).
\end{equation} 
Transforming to momentum space, the initial condition for the FRG flow is thus
\begin{align}
\label{eq:initcondfrglevel}
	V_0(\bm{k}_1,\bm{k}_2,\bm{k}_3,\bm{k}_4)=\frac{V_1}{2} \underset{\bm{\delta}}{\sum} \bigg(e^{-i(\bm{k}_2-\bm{k}_4) \bm{\delta}}-e^{-i(\bm{k}_2-\bm{k}_3)\bm{\delta}}-e^{-i(\bm{k}_1-\bm{k}_4)\bm{\delta}}+e^{-i(\bm{k}_1-\bm{k}_3)\bm{\delta}}\bigg) \,,
\end{align}
where we sum over the nearest-neighbor displacement vectors $\bm{\delta}$. Projecting all momenta to the Fermi surface via $\pi: 1. \text{BZ} \to \mathbb{Z}^{N}_{\text{FS}}$, Eq.~\eqref{eq:initcondfrglevel} directly serves as the initial condition for the patching scheme. 

For the TUFRG approach, we additionally insert Eq.~\eqref{eq:initcondfrglevel} into Eqs.~\eqref{eq::crossprojection0}-\eqref{eq::crossprojection} to derive explicit expressions for $V_{l,l^\prime}^{X,0}$, with $X\in\{P,C,D\}$. This procedure finally yields
\begin{align}
    V^{C,0}_{\mathbf{R}_1,\mathbf{R}_1}(\mathbf{q})=-V^{D,0}_{\mathbf{R}_1,\mathbf{R}_1}(\mathbf{q})&=-V_1\underset{\bm{\delta}}{\sum}e^{i\mathbf{q}\bm{\delta}}\\
    V^{P,0}_{\mathbf{R}_l,\mathbf{R}_l}(\mathbf{q})=V^{C,0}_{\mathbf{R}_l,\mathbf{R}_l}(\mathbf{q})=-V^{D,0}_{\mathbf{R}_l,\mathbf{R}_l}(\mathbf{q})&=V_1\\ V^{P,0}_{\mathbf{R}_{-l},\mathbf{R}_l}(\mathbf{q})&=-V_1 e^{-i\mathbf{q}\mathbf{R}_l} \,,
\end{align}
with $l \in \{2,3,4,5,6,7\}$ as the initial condition for the TUFRG flow.

\section{Chern numbers}
\label{app:chern}

To access possible topological properties of pairing instabilities, we consult a Skyrmion winding number formula~\cite{PhysRevB.61.10267,Black_Schaffer_2014}
\begin{align}
    \mathcal{C} = \frac{1}{4 \pi} \int_{1. \mathrm{BZ}} d^2 k \ \left \langle \bm{m}(\bm{k}) \bigg{|} \frac{\partial \bm{m}(\bm{k})}{\partial k_x} \times \frac{\partial \bm{m}(\bm{k})}{\partial k_y} \right \rangle \,,
\end{align}
where $\langle . \ | \ . \rangle$ is the Euclidean scalar product. Here, $\bm{m}(\bm{k})$ denotes the pseudospin vector or Skyrmion magnetization, which follows the winding of the superconducting gap around the Fermi surface. In algebraic form,  $\bm{m}(\bm{k})$ is given by
\begin{align}
    \bm{m}(\bm{k}) = \frac{1}{E(\bm{k})} 
    \begin{pmatrix} 
        \mathrm{Re}(\Delta(\bm{k})) \\ 
        \mathrm{Im}(\Delta(\bm{k})) \\ 
        \xi(\bm{k})
    \end{pmatrix} \,,
\end{align}
where $E(\bm{k}) = \sqrt{|\Delta(\bm{k})|^{2} + \xi(\bm{k})^2}$ is the Bogoliubov quasi-particle spectrum. 

\begin{figure}[t!]
    \centering
    \includegraphics[width=0.5\columnwidth]{ 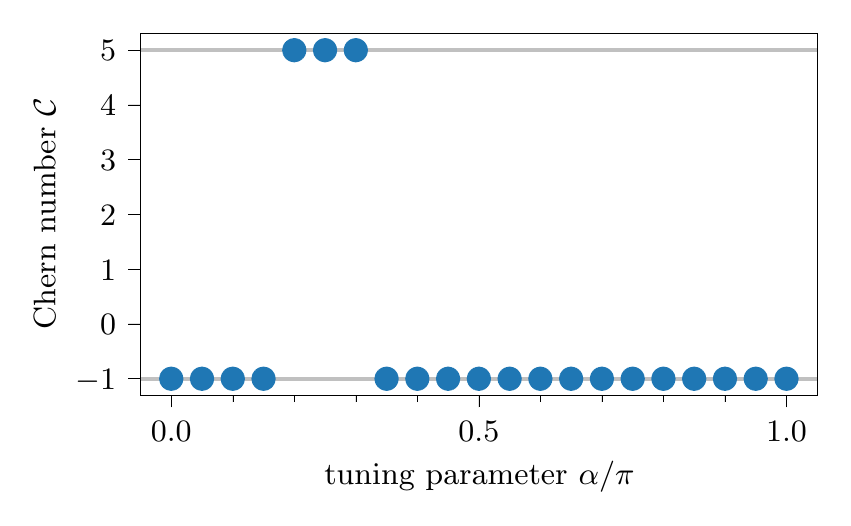}
    \caption{\textbf{Example calculations for Chern numbers in the $E_1$ representation for $\mu / t = 1.7$.} Motivated by our finding of a higher-harmonic $\tilde{p}$-wave instability for repulsive interactions $V_1 / t > 0$ (see Figs.~\ref{fig:repulsive17_patching} \& \ref{fig:repulsivepanel17}), we perform exemplary computations of $\mathcal{C}$ for superconducting gaps of the form $\Delta(\bm{k}) = [\text{cos}(\alpha) \delta^{E_1}_{1}(\bm{k}) + \text{sin}(\alpha)\tilde{\delta}^{E_1}_{1}(\bm{k})] + i \times [\text{cos}(\alpha) \delta^{E_1}_{2}(\bm{k}) + \text{sin}(\alpha)\tilde{\delta}^{E_1}_{2}(\bm{k})]$, where $\delta^{E_1}_{1(2)}$ denotes the nearest-neighbor lattice harmonics of the $E_1$ irrep. and $\tilde{\delta}^{E_1}_{1(2)}$ the respective second neighbor functions. The model is chosen such that we recover the pure first (second) neighbor limit for $\alpha = 0 \ (\pi)$.}
    \label{fig:Chern}
\end{figure}

It is immediately clear, that any real or purely imaginary gap function will result in a topologically trivial state with $\mathcal{C} = 0$. In contrast, for a gap function corresponding to a two-dimensional irreducible representation, such as the $p$-wave instabilities we found in the main text, the possibility of non-trivial topology arises. In principle, one would need to minimize the mean-field free energy for a linear superposition of the respective lattice harmonics and determine whether or not a complex gap function prevails. Here, we resign from employing this variational approach and instead use a heuristic argument. Consider the ground state energy $E_{0} = -\langle |\Delta(\bm{k})| \rangle_{\mathrm{FS}}$ for a gap function $\Delta(\bm{k})$ which we suppose to live in the complex two-dimensional space corresponding to a doubly degenerate eigenvalue of the linearized gap equation. If this linear combination is either real or purely imaginary, there will be momenta on the Fermi surface where $|\Delta(\bm{k})|$ is gapless and no contribution to the ground state energy is obtained henceforth. If one assumes a complex linear combination instead, $|\Delta(\bm{k})|$ will be fully gapped at the Fermi level and thus, a lower ground state energy is obtained. It is therefore natural to assume, that the energetically more beneficial superposition of lattice harmonics is a complex one. Computing $\mathcal{C}$ from the ansatz $\Delta(\bm{k}) = \delta^{E_1}_{1}(\bm{k}) + i \delta^{E_1}_{2}(\bm{k})$ for the nearest-neighbor or second neighbor lattice harmonics $\delta^{E_1}$ of the $E_1$ irrep., for example, we find $\mathcal{C} = -1$ over the entire range of fillings where the $p$-wave instability occurs. An admixture of both, the first and second neighbor functions may, however, yield a strongly enhanced Chern number, as exemplified in Fig.~\ref{fig:Chern}.

\end{document}